\documentclass[aps]{revtex4}
\usepackage{graphicx}
\begin{document}
\draft
\title{Superfluid equation of state of cold fermionic gases in the Bose-Einstein regime}
\author{R. Combescot$^{(a),(b)}$ and  X. Leyronas$^{(a)}$}
\address{(a) Laboratoire de Physique Statistique,
 Ecole Normale Sup\'erieure,
24 rue Lhomond, 75231 Paris Cedex 05, France}
\address{(b) Institut Universitaire de France}
\date{Received \today}

\begin{abstract}
We present an exact many-body theory of ultracold fermionic gases for the Bose-Einstein condensation (BEC) regime
of the BEC-BCS crossover. This is a purely fermionic approach which treats explicitely and systematically
the dimers formed in the BEC regime as made of two fermions. We consider specifically the zero temperature case and
calculate the first terms of the expansion of the chemical potential in powers of the density $n$.
We derive first the mean-field contribution, which has the expected standard expression 
when it is written in terms of the dimer-dimer scattering length $a_M$. We go next in the expansion to
the Lee-Huang-Yang order, proportional to $n^{3/2}$. We find the far less obvious result that it retains also
the same expression in terms of $a_M$ as for elementary bosons. The composite nature of the dimers appears  
only in the next term proportional to $n^2$.
\end{abstract}
\maketitle

\section{INTRODUCTION}

The recent experimental developments in ultracold Fermi gases and in particular the experimental 
realization of the BEC-BCS crossover through Feshbach resonance
\cite{gps} have arised in a  field where theoretical investigations started a long time ago. Indeed fairly soon after 
the publication of the Bardeen-Cooper-Shrieffer (BCS) theory \cite{bcs}, the extension to the neutral Fermi liquid $^3$He was considered \cite{pitcoop}.
Although the case of liquid $^3$He is more complex than the one of ultracold gases, it is nevertheless analogous to the situation found on the BCS side of the crossover, in the strong interaction regime.
Similarly the possibility of Bose-Einstein condensation (BEC) of composite bosons was considered quite early for excitons in
semiconductors \cite{mosblat}, where the composite nature of these bosons is expected to be an essential physical feature.
This is the situation met in the BEC range of the BEC-BCS crossover.

Deep theoretical investigation of this last situation was made not long afterwards by Keldysh and Kozlov \cite{kk}, following a
seminal work by Popov \cite{popov} which considered a physical model with short range potentials, much closer to the
ultracold gas case. In both cases, due to the formal similarity between the Bogoliubov and the BCS frameworks, the theoretical
treatment relied on the fact that in the dilute limit the formalism leads directly to the Schr\"odinger equation for the molecular state
corresponding to the composite boson. This provided an anticipation of the study of the whole BEC-BCS crossover. 
Independently Eagles \cite{eagles}, in the course of a study of superconductivity in doped semiconductors 
(where the electron gas has a low concentration compared to standard metals), 
was led to investigate the extrapolation of the simple BCS formalism toward the dilute
regime where pair formation would take place, due to the attractive interaction,
and Bose condensation of these pairs should occur at lower temperatures.

In the context of cold gases, the physics of the crossover was considered  by Leggett \cite{leg},        who stressed the physical
interest of considering Cooper pairs as giant molecules, emphasized the change of physical regime when the chemical
potential goes through zero, and introduced the scattering length in the BCS formulation. The matter was taken up by
Nozi\`eres and Schmitt-Rink \cite{nsr} who studied the smooth evolution of the critical temperature in the crossover within
the simplest $T-$matrix approximation, showing in particular how the critical temperature of the ideal Bose gas is recovered
in the strong coupling limit. S\'a de Melo, Randeria and Engelbrecht \cite{sdm} addressed the crossover with the functional
integral formalism.

As it has often been stressed, the physical situation found in this BEC-BCS crossover is extremely interesting
since (at least for wide Feshbach resonances \cite{gps,rcfesh}) the scattering length $a$ of the different fermions with
mass $m$ is enough to fully describe
their interaction, while one has still to deal with the full complexity of a strongly interacting system. As a result the
theoretical problem has not been solved exactly and one has to rely on various approximate schemes \cite{gps},
the simplest one being the straight BCS formalism (often called BCS-MF) which is known to give a correct description
at zero temperature in both the
extreme BEC and BCS limits $a \rightarrow 0_{\pm} $. The imperfections of the approximate treatments are clearer
in the BEC regime. Indeed if one wants to describe properly the departure from the ideal Bose gas of dimers, one
has to provide a correct description of the dimer-dimer interaction, in order to find the expected physics of a weakly
interacting Bose gas. As stressed by Keldysh and Kozlov \cite{kk} the composite nature of the dimers has already
to be properly taken into account at this stage in order to obtain correct  results. 

A first step in this direction for the specific case of short range interactions, relevant for cold gases, was made by Haussmann \cite{haus} who obtained,
both in the normal and in the superfluid state (where this can be found naturally out of the BCS ansatz),
that the dimer-dimer
scattering length $a_M$ is related to the fermion scattering length $a$ by $a_M=2a$. This result was also obtained
in Ref.\cite{sdm} in the superfluid state by other methods. The level of approximation corresponding to this result turns
out to be the Born approximation for the dimer-dimer scattering. At the same level of approximation Pieri and Strinati \cite{pieristrinati} derived quite recently the Gross-Pitaevskii equation from the 
Bogoliubov-de Gennes equations. However this level of approximation was markedly improved by Pieri and Strinati
\cite{pierist} who performed the calculation at the level of a $T-$matrix approximation and found $a_M \simeq 0.75\,a$.
Nevertheless their calculation does not include all the processes resulting from the existence of the fermions making
up these composite bosons. The exact numerical result was obtained recently by Petrov, Salomon and Shlyapnikov
\cite{petrov}
who basically solved the relevant Schr\"odinger equation for the four-body problem and found for this dimer-dimer
scattering length $a_M \simeq 0.60\,a$. Thereafter it was shown \cite{bkkcl} how this same problem could be exactly
formulated and solved within the methods of quantum field theory, with naturally exactly the same result. This last
work is quite relevant to the present paper, and represents the basis out of which we will work. This is naturally linked
to the fact that field  theoretic methods are a convenient, and probably necessary, framework for an exact formulation
and solution of this superfluid strong coupling problem raised by the BEC-BCS crossover.

In this paper, we provide an exact approach to the BEC-BCS crossover problem from the BEC side
by presenting an exact fermionic theory 
of a BEC superfluid of composite bosons in the low density range. The two interacting fermion species we deal with
are actually the two lowest energy hyperfine states used experimentally \cite{gps} in ultracold gases of $^6$Li or $^{40}$K, 
and as it is usually done we will for convenience refer to them as "spin up" and "spin down" states.
Specifically working in the low density range means that we proceed basically to
an expansion in powers of the gas density. Otherwise our framework is completely general. In the present paper
we deal with the lower
orders in this expansion, but nothing in principle prevents our approach to be extended to higher orders, although
admittedly this will require much more work. Similarly we will restrict ourselves to the $T=0$ situation, which brings strong simplifications as we will see because in this case there are no fermions at all in the normal state. However it is quite possible
to extend our approach to non zero temperature. Finally we deal here with thermodynamics, but our framework allows us
naturally to calculate dynamical quantities.

As we mentionned our approach is in principle a low density expansion. However the density $n$ is not a convenient
basic parameter since it rather appears in the formalism as the result of a calculation. On the other hand it is known
(and we will see it explicitely below)
that in the simple BCS approximation for the BEC limit, which as mentionned above corresponds to a  Born approximation
at the level of the mean field term, the density $n$ is proportional to $\Delta^2$, where $\Delta $ is the gap parameter of
the BCS theory. Hence a convenient way to proceed effectively to a low density expansion is rather to expand in powers
of $\Delta $. However this quantity is specific of the BCS approximation, which has no general validity. But it is directly
related to the anomalous self-energy $\Delta (p)$ (where $p$ is an energy-momentum four-vector) which is a completely
general microscopic quantity characteristic of a superfluid system. Hence our method will specifically be an expansion
of the general equations in powers of the anomalous self-energy $\Delta (p)$.

Here we will calculate the equation of state, or more precisely the dependence of the chemical potential $\mu $ on the density $n$
(a short account of our calculation has already been published \cite{xlrc}).
Actually this is rather the reciprocal of this function which comes naturally out of the calculation, since $\mu $ enters the
Green's functions while $n$ is obtained at the end of the calculation, after elimination of $\Delta (p)$ which comes in the
intermediate steps. The lowest order term in the expression of $\mu $, beyond the trivial term of half the molecular binding energy $-1/(2ma^2)$,
is the mean-field term. One has to note that, even at this stage, the composite nature of the bosons we deal with is entering
as noted by Keldysh and Kozlov \cite{kk}. However in our case it is lumped into the full dimer-dimer scattering length $a_M$.
This result is physically natural since, beyond the zeroth order term $-1/(2ma^2)$ in the chemical potential \cite{hbar} corresponding
to an isolated molecule, we expect in a density expansion to find the physics of two coexisting molecules, which should be fully
describable by $a_M$ for its static properties, corresponding to the zero energy scattering properties. Nevertheless, as far as we know, this quite reasonable result has always been in the literature more or less taken for granted or cursorily obtained, 
but not fully explicitely derived from a microscopic theory for composite bosons. 
Hence our first step will be to obtain this fully explicit derivation.

However we are naturally essentially interested by the next order term, at which we will also limit our present calculation.
From the theory of weakly interacting elementary bosons this term is proportional to $n^{3/2}$, as first obtained by
Lee and Yang \cite{lhy}. We refer to this term as the LHY term. This term has then been rederived by Beliaev \cite{bel} making use
of field theoretic methods. Naturally it is reasonable to expect that a similar term arises for composite bosons. 
What is however less clear is that the coefficient is just the same as for elementary bosons, 
i.e. that it can be simply expressed in the same way in terms of $a_M$.
Indeed it is clear that, at some stage in the expansion, the composite nature of the bosons will enter by other quantities
than those describing the physics of elementary bosons. This is quite obvious physically since, by going to higher densities
(which corresponds to take into account higher order terms in the expansion), we will go toward unitarity and find physical
properties linked to the existence of Fermi seas. Obviously this can not be obtained from the physics of elementary bosons.
Hence it is conceivable that additional processes contribute to the  $n^{3/2}$ term for composite bosons.
Nevertheless we will find at the end of our paper that the $n^{3/2}$ term is indeed identical to the LHY term 
provided the dimer-dimer scattering length is used. The fact that this is not obvious is immediately realized from the
number and the complexity of our steps, which do not map systematically on the elementary boson derivation.

On the other hand this result makes sense physically when it is realized that the LHY term is directly
linked to the existence of the gapless collective mode. A standard perturbation expansion should produce
a $n^2$ term as the next term after the mean-field one. The existence of the LHY term is linked to the presence
of these gapless elementary excitations which produce a singularity in the expansion, leading formally to
a divergence in the coefficient of the $n^2$ term. Obtaining the proper result is equivalent to the resummation of a
partial series to all order in perturbations in order to get rid of the singularity. Since these steps involve only low
energies, they do not test the internal structure of the bosons which comes into play when energies of order of
the binding energy $1/ma^2$ are involved. In such a case it is reasonable that no difference appears between
elementary and composite bosons. On the other hand the calculation of the coefficient of the regular $n^2$ term
in the expansion involves the consideration of energies of order of the binding energy, as we will see explicitely, 
making explicit the composite nature of the bosons. This implies that at the level of this term one finds a result different
from the case of elementary bosons. Note that this result has been previously assumed to be correct
in calculations of collective mode frequencies \cite{sandro,acls}. On one hand this was in agreement with Monte-Carlo 
calculations \cite{abcg}, and on the other hand this has been supported by very recent experiments \cite{altm}.

At the level of handling the collective mode,
very strong analogies with a purely bosonic approach, in particular with Beliaev's work \cite{bel}
will appear for obvious physical reasons, although it can be seen that our treatment
bypasses a good deal of Beliaev's detailed diagrammatic analysis.
We will not try to put these analogies away, but rather make use of these similarities and correspondances to make 
easier the  physical and technical understanding of our calculation. However we will never make an actual 
use of a bosonic approach and all our theory is a purely fermionic theory,
the essential physical feature being that we deal with fermion physics but with negative chemical potential $\mu <0$. 
In particular we often use the word dimer to describe two fermions with spin up and spin down. 
But since the theory contains, in the corresponding propagator, not only the bound state of these fermions 
but also the scattering states, it should be taken as a simple convenient wording, although it is naturally physically suggestive.
More generally and formally the anomalous self-energies $\Delta(p) $ and $\Delta ^*(p)$ play clearly a role analogous to 
$\langle b_0 \rangle$ and $\langle b_0^{\dag}\rangle$ for bosonic superfluidity.

Our paper is organized according to the above introduction. After introducing briefly the general formalism
in the next section \ref{general}, we proceed to the calculation of the normal and anomalous self-energies in section \ref{deltexp}
to lowest order in $\Delta (p)$, which leads us to the mean field term in the expression of the chemical potential.
Then in section \ref{collmod} we proceed to discuss the collective mode which plays an essential role in the calculation
of the LHY term. Then the calculation of the contribution of this collective mode to the normal and anomalous self-energies
is performed in section \ref{sigcollmod}, at the end of which the LHY contribution to the chemical potential is obtained.
A number of technical details are treated in appendices, to try to make more palatable the main part of this paper
which is already fairly heavy. 

\section{general}\label{general}

Our work deals only with the $T=0$ situation. Hence it would be most natural to use the standard $T=0$ Green's function formalism and naturally all the content of our paper can be reproduced with this formalism. However one has in this case to be careful in dealing with real frequencies, and specify precisely the location, with respect to the real axis, of singularities in the corresponding complex plane. This is often left implicit. A simple way to clarify this question is to deal with complex frequencies. But this is actually what one has to do naturally when one works at $T \neq 0$, where the frequencies run on the imaginary frequency axis. In the $T \rightarrow 0$ limit the discrete sums over Matsubara frequencies turn into integrals over frequencies running on the imaginary axis (the link with the $T=0$ formalism is merely obtained by continuing analytically the frequencies toward the real frequency axis, that is performing a so-called Wick rotation on the frequency). This leads to somewhat clearer results. Hence, despite its somewhat artificial character, we will work quite specifically with this $T=0$ limit of the $T \neq 0$ formalism. Actually this will not appear in many our expressions because of their formal nature and for this reason it does not make any real problem. The essential point is that, in this way, we will always have perfectly regular expressions, with an effective handling which is quite clear and easy.

We deal only with the formalism of fermionic superfluidity, as it is known in superconductivity \cite{agd}
and superfluid $^3$He \cite{vw}. This entails the use of both the 'normal' fermionic Green's functions $G_{\uparrow \uparrow }(p)= G_{\downarrow \downarrow }(p) \equiv G(p)$, where we make use of the four-vector notation $p \equiv \{{\bf p},\nu \}$, and their 'anomalous' counterpart $F(p)$ and $F^{+}(p)$, where we make use of standard notations \cite{agd}. 
This could be written in a compact Nambu formalism, but for clarity we prefer to write the normal and anomalous parts explicitely,
with slightly different sign conventions.
In practice we will always consider in the following, when we calculate $G(p)$, that we deal with $G_{\uparrow \uparrow }(p)$, without writing it explicitely for simplicity. The existence (i.e. the fact that they are non zero) of these
anomalous Green's function is the basic ingredient of fermionic superfluidity in our formalism. This is quite clear physically since they are related to the condensate.

\begin{figure}
\centering
\includegraphics[width=100mm]{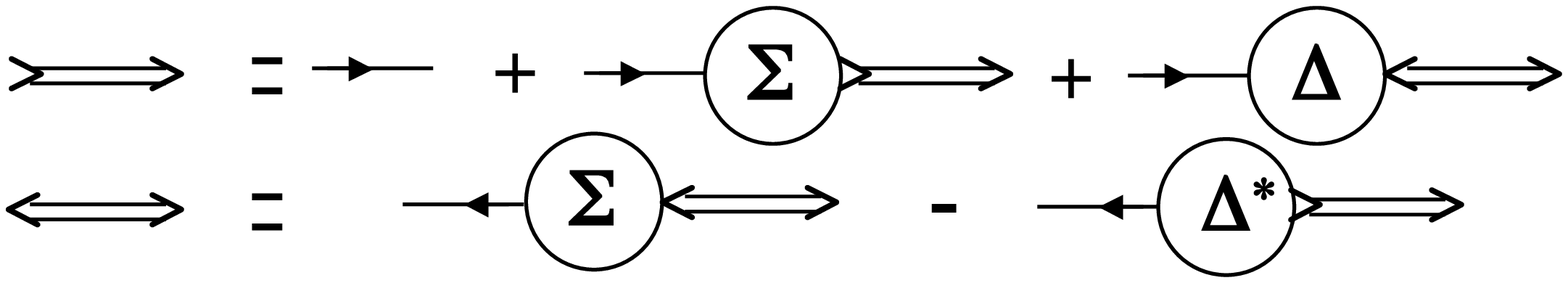}
\caption{Diagrammatic representation of Dyson's equations Eq.(\ref{eqdyson},\ref{eqdysona})}
\label{FigDyson}
\end{figure}

We start by writing Dyson's equations for this superfluid:
\begin{eqnarray}\label{eqdyson}
G(p)&=&G_0(p)+G_0(p) \Sigma (p) G(p) +G_0(p) \Delta (p) F^{+}(p) \\
F^{+}(p)&=&G_0(-p) \Sigma (-p) F^{+}(p) -G_0(-p) \Delta^{*} (p) G(p) \label{eqdysona}
\end{eqnarray}
their corresponding diagrammatic representation being given in Fig. \ref{FigDyson}. Here $G_0(p)=
[i\nu -\xi_{\bf p} ]^{-1}$ is the free atom Green's function, with $\xi_{\bf p}=\epsilon _{\bf p}-\mu $ the atom kinetic energy $\epsilon _{\bf p} = {\bf p}^2/2m$ measured from the chemical potential $\mu $. The quantities $\Sigma (p)$ and $\Delta (p)$ are respectively the normal and the anomalous self-energies. By definition $\Sigma (p)$ is the sum of all diagrams, with one entering $p$ line and one outgoing $p$ line, which can not separated into two disconnected parts by cutting a single $G_0(p)$ line (or $G_0(-p)$ line). $\Delta (p)$ has the same definition except that it has two entering lines, one ($p, \uparrow$) line and one ($-p,\downarrow$) line.
We note that $\Delta (p)$ depends naturally in the general case on momentum and frequency. In the second equation, for $F^{+}(p)$ which begins with an outgoing ($-p,\downarrow$) line and finishes with an outgoing ($p, \uparrow$) line, it would look more symmetrical to introduce a corresponding anomalous self-energy $\Delta ^{+}(p)$, analogous to $\Delta (p)$, but with two outgoing lines, one ($-p, \downarrow$) line and one ($p,\uparrow$) line. However it can be related directly to $\Delta^{*} (p)$ by time reversal, as we have written explicitely. All formulae in the following turn out to be simpler to interpret when this substitution is done, and we will follow this standard habit. We note also that, since we deal with an equilibrium situation, where the gas is homogeneous, we do not need to introduce Green's functions with different entering and outgoing momenta, as it would be required in the case of an inhomogeneous system.

Dyson's equations Eq.~(\ref{eqdyson}) can be understood as a way of ordering the full perturbation expansion, by gathering terms into $\Sigma (p)$ and $\Delta (p)$. However it is interesting to note that this can be done in slightly different ways. For example we can gather in Eq.~(\ref{eqdyson}) all the terms where only the free Green's function $G_0(p)$ and the normal self-energy $\Sigma (p)$ appear, that is the terms where no anomalous terms are present. This leads to introduce ${\mathcal G}_0(p)= G_0(p)+G_0(p)\Sigma (p)G_0(p)+G_0(p)\Sigma (p)G_0(p)\Sigma (p)G_0(p)+...$, that is ${\mathcal G}^{-1}_0(p)= G^{-1}_0(p)-\Sigma (p)$. In this way, by making use of ${\mathcal G}_0(p)$ instead of $G_0(p)$, we are only left to write in Dyson's equations the anomalous terms, that is to rewrite these equations as:
\begin{eqnarray}\label{eqdyson1}
G(p)={\mathcal G}_0(p)+{\mathcal G}_0(p) \Delta (p) F^{+}(p)  \\
F^{+}(p)= -{\mathcal G}_0(-p) \Delta^{*} (p) G(p) \label{eqdyson1a}
\end{eqnarray}
It is indeed checked easily that the algebraic solution of Eq.~(\ref{eqdyson},\ref{eqdysona}), as well as the one of Eq.~(\ref{eqdyson1},
\ref{eqdyson1a}), is indeed:
\begin{eqnarray}\label{eqdysonsol}
G^{-1}(p)={\mathcal G}^{-1}_0(p)+\Delta (p) {\mathcal G}_0(-p) \Delta^{*} (p)
\end{eqnarray}
Making use of the expression of ${\mathcal G}^{-1}_0(p)$ this can be rewritten as:
\begin{eqnarray}\label{eqdysonsol1}
G^{-1}(p)=G^{-1}_0(p)-\left[\Sigma (p)-\Delta (p) {\mathcal G}_0(-p) \Delta^{*} (p)\right]
\end{eqnarray}
This result could have been obtained directly in the following way. In the definition we have taken above
for the self-energy, we had excluded diagrams made of two blocks linked by a single propagator $G_0(p)$ (corresponding to an atom propagating to the right, as for the definition of the normal self-energy) or $G_0(-p)$ (corresponding to an atom propagating to the left, as it is produced by the anomalous self-energies). However we could take another definition where we would exclude only blocks linked by a rightward propagator, just as in the normal state. In this case the formal expression of $G(p)$ in terms of the self-energy, i.e. $G^{-1}_0(p)= G^{-1}_0(p)-{\mathcal S} (p)$, is just the same as in the normal state, as it is indeed the case in Eq.~(\ref{eqdysonsol1}). But the expression of the self-energy has been naturally modified from $\Sigma (p)$ to ${\mathcal S} (p)$. Indeed we have to include now terms containing leftward propagators $G_0(-p)$. But it is easily realized that these new terms contain at most a product $\Delta (p) \Delta^{*} (p)$, corresponding to the presence of anomalous self-energies of opposite types at the extremities of the block. These two anomalous self-energies can be separated by
any number of leftward propagators $G_0(-p)$, with normal self-energies $\Sigma (-p) $. The sum of all these produces a factor ${\mathcal G}_0(-p)$, which leads to the relation:
\begin{eqnarray}\label{eqsigm}
{\mathcal S} (p) = \Sigma (p)-\Delta (p) {\mathcal G}_0(-p) \Delta^{*} (p)
\end{eqnarray}
in agreeement with Eq.~(\ref{eqdysonsol1}).

\section{$\Delta$ expansion: lowest order}\label{deltexp}

While they are completely general, the considerations presented in the preceding section are not of much practical use since we have no expressions for the normal and anomalous self-energies. In order to obtain explicit expressions, we will now  begin to proceed along the lines outlined in the introduction, namely to perform a systematic expansion in powers of $\Delta $. In the present section, as an introduction and as a first step, we will only consider the lowest order in this expansion. The next order will then be addressed in the next section. In the first subsection we will consider the expansion of the normal self-energy, which turns out to be the easier one. Then the anomalous self-energy will be handled along the same lines.

\subsection{Normal self-energy}\label{normself}
\begin{figure}
\centering
\includegraphics[width=60mm]{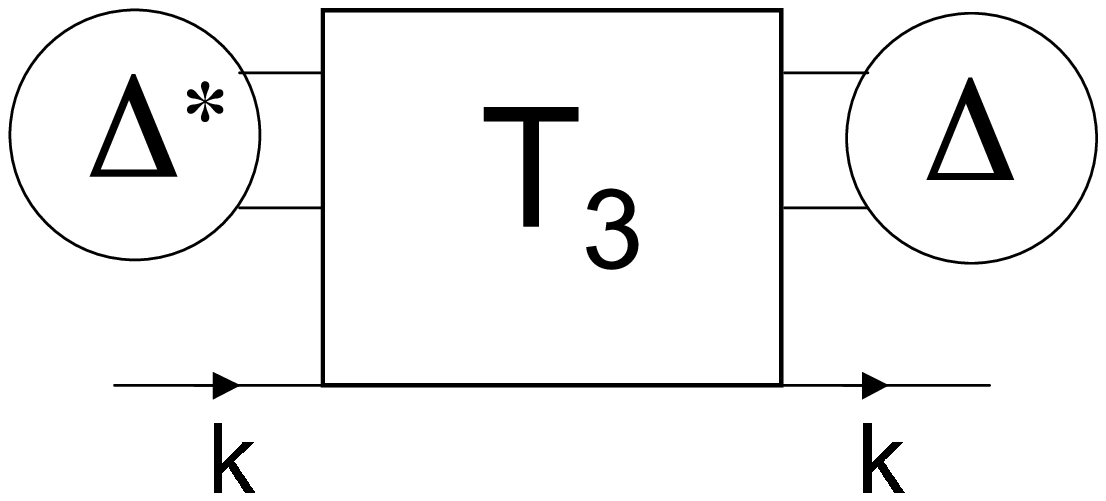}
\caption{Diagrammatic representation of ${\mathcal S}^{(2)}(p)$}
\label{FigT3}
\end{figure}

As we have indicated in the introduction, and as we will check explicitely below, the expansion in powers of $\Delta $ has the character of an expansion in powers of the density $n$ since we will find
$n \sim \Delta ^2$. Hence the zeroth order in powers of $\Delta $ corresponds to an infinitely dilute gas
for which the normal self-energy is naturally zero. Since by its definition this normal self-energy contains only processes which conserve globally the number of particles, the lowest order term $\Sigma^{(2)}(p)$ in the normal self-energy must contain both $\Delta $ and $\Delta ^{*}$, because $\Delta $ implies from its definition the annihilation of two atoms (going physically in the condensate), while $\Delta ^{*}$ corresponds to the creation of two atoms (coming out of the condensate). Hence this lowest order term in the normal self-energy is the sum of all diagrams which contain $\Delta (p')$ and $\Delta ^{*}(p'')$, and which have one entering line $p$ and one outgoing line $p$, the rest of the diagrams containing only (since we proceed to an expansion) $\Delta = 0$ quantities, that is normal state quantities (see Fig. \ref{FigT3}). We call the sum of all these normal state diagrams the 'normal part' of $\Sigma^{(2)}(p)$. This is just $\Sigma^{(2)}(p)$, except for the $\Delta (p')$ and $\Delta ^{*}(p'')$ terms. 

Explicitely this means that this 'normal part' of the diagrams contains only any number of bare propagators $G_0(p)$ and of interactions lines. Moreover the two atom lines coming out of $\Delta ^{*}(p'')$(with opposite spins) are not allowed to interact immediately after entering the 'normal part' of the diagram, since such an interaction is already included in $\Delta ^{*}(p'')$ (as we will see explicitely below) and allowing it would amount to double counting of diagrams. Similarly the two atom lines entering $\Delta (p')$ are not allowed to interact immediately before. As it happens, such a normal state set of diagrams has already been essentially considered in Ref.\cite{bkkcl}. Indeed in this paper the scattering of a single atom by a dimer (made of atoms with opposite spins) has been calculated. The fact that a dimer is entering the diagram implied that the two corresponding atom lines can not interact immediately in the diagram, since such an interaction is already taken into account in the dimer propagator and this would be again double counting. There is a similar requirement for the two lines corresponding to the outgoing dimer. We see that the requirements on the normal part of
$\Sigma^{(2)}(p)$ are just the same as in the above paper. In this Ref.\cite{bkkcl} the sum of all these diagrams was denoted $T_3(p_1,p_2;P)$ where $p_1$ (respectively $p_2$) is the four-vector of the entering (respectively outgoing) atom, and $P$ is the total four-vector of the single atom and the dimer.
In our statement we apparently overlook the fact that $T_3(p_1,p_2;P)$ contains only three atom lines, which go from the beginning to the end of the diagram, with various interactions between them in the diagram, since we deal with the vacuum scattering of an atom and a dimer. In contrast we could think that, in addition to these lines, the 'normal part' of $\Sigma^{(2)}(p)$ may contain any number of loops of atom lines. However it is clear physically that these loops are not allowed. Indeed if we think of this 'normal part' as representing, in a time representation, a succession of processes due to interaction, the appearance of a loop corresponds to the creation of a particle-hole pair. While such processes can always occur at non zero temperature, in our $T=0$ situation they are possible only if the chemical potential $\mu $ is positive, thereby allowing the existence of a Fermi sea in which the hole can be created. But in our dilute regime we have $\mu  < 0$ and these processes are forbidden. This argument is developped more technically in Appendix \ref{appnoloop}. Hence it is correct to identify the 'normal part' of $\Sigma^{(2)}(p)$ with $T_3$ since both contain only three atom lines, subject to the same restrictions indicated above.

Nevertheless there is still a small difference in the general case between the 'normal part' we want and $T_3$. In the definition of $T_3$ one has to sum over all possible four-vectors for the two lines making up the entering dimer, with the only restriction that their sum is fixed at $P-p_1$. A similar summation applies for the two lines making up the outgoing dimer. On the other hand in our 'normal part' the two lines (p',$\uparrow$) and (-p',$\downarrow$) entering in $\Delta (p')$ have a total zero four-vector, and because $\Delta (p')$ depends on $p'$, when we sum over $p'$, there is a weight $\Delta (p')$ which depends on $p'$, in contrast with the case of $T_3$. Nevertheless, as shown in Appendix \ref{appT3}, it is possible to express in the general case our self-energy in terms of $T_3$. But in our specific problem, this is an unnecessary complication since we will see that, at the order of the expansion we are considering, the dependence of  $\Delta (p')$ on $p'$ can be neglected, and $\Delta (p')$ can be considered as a constant which we denote merely as $\Delta $. Hence we can factor $\Delta$ out. Similar considerations apply to $\Delta ^{*}(p")$  and the entering dimer lines. 

There is still a very important point to consider before writing our expression for $\Sigma^{(2)}(p)$. Among all the diagrams we have considered contributing to $\Sigma^{(2)}(p)$, there is necessarily a diagram which gives a contribution $-\Delta G_0(-p) \Delta ^{*}$ because it satisfies all our requirements. Indeed this is just what is coming from the lowest order diagram in $T_3$, which gives a contribution $-G_0(-p)$ to $T_3$ and corresponds merely to the Born approximation. This diagram is necessarily present because we considered it already in the preceding section \ref{general}, when we were dealing with the general expansion of $G(p)$ in powers of $\Sigma(p)$ and $\Delta (p)$. This is just the last term in Eq.(\ref{eqdysonsol1}), when we replace ${\mathcal G}_0(-p)$ by $G_0(-p)$ (since we want only 'normal state' diagrams) and take $\Delta (p)$ independent of $p$. However, since it is already explicitely present in Eq.(\ref{eqdysonsol1}), this implies automatically that it is reducible (and indeed it is obviously reducible) and for this reason should not be included in $\Sigma^{(2)}(p)$ (otherwise we would double count this diagram). On the other hand this is clearly the only reducible contribution coming from $T_3$, as seen directly from the diagrammatic expansion \cite{bkkcl} of $T_3$, or because any such reducible contribution has to appear explicitely from Eq.(\ref{eqdysonsol1}) as explained above. Accordingly we have to subtract this contribution out and we obtain finally:
\begin{eqnarray}\label{eqsigmn2}
\Sigma^{(2)}(p)=|\Delta |^2\,T_3(p,p;p)+|\Delta |^2\, G_0(-p)
\end{eqnarray}
We can also rewrite this result in terms of our self-energy ${\mathcal S}$ as:
\begin{eqnarray}\label{eqsigmn2bis}
{\mathcal S}^{(2)}(p)=|\Delta |^2\,T_3(p,p;p)
\end{eqnarray}
This result is also directly quite natural since, when we evaluate ${\mathcal S}$, the diagrams with $G_0(-p)$ are not considered as reducible, as we have seen, and accordingly $-\Delta G_0(-p) \Delta ^{*}$ should not be subtracted.

We are now in position to calculate the single spin atomic density $n$, which is given quite generally by:
\begin{eqnarray}
n=\int \frac{d{\bf p}}{(2\pi )^3}\int_{-\infty}^{+\infty}\frac{d\nu}{2\pi} \;G(p) \,e^{i\nu 0_+}
\label{eqdefn}
\end{eqnarray}
If $G$ is replaced by $G_0$, one checks easily that $n=0$, which is physically obvious since in this case we deal with a non interacting fermion gas, at $T=0$, and chemical potential $\mu <0$. There is no Fermi sea (which arises only for $\mu >0$) and there are no fermions. Hence we can replace completely generally $G(p)$ by $G(p)-G_0(p)=G_0(p){\mathcal S}(p)G(p)$ in Eq.~(\ref{eqdefn}). Since we work in this section at the lowest order in $\Delta $, which means at order $\Delta ^2$ in the present case, we can replace ${\mathcal S}(p)$ by ${\mathcal S}^{(2)}(p)$, and $G(p)$ by $G_0(p)$. This leads to:
\begin{eqnarray}
n=|\Delta |^2\,\int \frac{d^3 {\bf p}}{(2\pi )^3}\int_{-\infty}^{+\infty}\frac{d\nu}{2\pi}\; T_3(p,p;p)\, \left[G_0(p)\right]^2
\label{eqdefn1}
\end{eqnarray}
where, as in Eq.~(\ref{eqdefn}), $p$ is for $\{{\bf p},\nu\}$.

At this stage it seems that, in order to calculate the density, we are required to have a deep knowledge of  $T_3$. This seems physically reasonable since, in Eq.~(\ref{eqdefn1}), $T_3$ describes the scattering of an atom with the condensate which has clearly to be taken into account. However the actual situation turns out to be much simpler. In order to calculate the integral over $\nu $ in Eq.~(\ref{eqdefn1}), it is convenient to close the contour in the upper complex half-plane ${\mathrm Im}\,\nu >0$ (which corresponds to ${\mathrm Re}\,(i\nu) <0$), since the pole of $G_0(p)=[i\nu -\xi_{\bf p}]^{-1}$ is found for ${\mathrm Im}\,\nu <0$ (since $\xi_{\bf p} = \epsilon _{\bf p} + |\mu |>0$). However it is shown in Appendix \ref{appT3anal} that, except for the Born contribution, $T_3(p,p;p)$ is analytical in the upper complex half-plane and does not contribute to the integral. The result comes entirely from the Born contribution. This means that surprisingly the contribution of $\Sigma^{(2)}(p)$ to $n$ is zero, and that the only non zero contribution comes from the explicit BCS term $-|\Delta|^2 G_0(-p)$ in Eq.~(\ref{eqsigm}). In other words the approximate BCS formalism gives the exact result at the lowest order in our $\Delta $ expansion. It reads:
\begin{eqnarray}\label{eqdefn2}
n=-|\Delta |^2\,\int \frac{d^3 {\bf p}}{(2\pi )^3}\int_{-\infty}^{+\infty}\frac{d\nu}{2\pi}\; G_0(-p)\, \left[G_0(p)\right]^2=\frac{m^2\,|\Delta |^2}{8\pi (2m|\mu |)^{1/2}}
\end{eqnarray}
If we make use in this result of the zeroth order value of the fermion chemical potential $\mu =-E_b/2=-1/2ma^2$, we end up with:
\begin{eqnarray}\label{eqdefn3}
n\simeq \frac{m^2\,a\,|\Delta |^2}{8\pi}
\end{eqnarray}
Actually this relation $|\Delta |^2=(8\pi /m^2 a)\,n$ is precisely obtained from the dimer propagator $T_2(P)$ (see section
\ref{collmod}) when it is evaluated near its pole and with the purely bosonic propagator factor merely replaced by $n$.   

\subsection{Anomalous self-energy}\label{anomself}

\begin{figure}
\centering
\includegraphics[width=120mm]{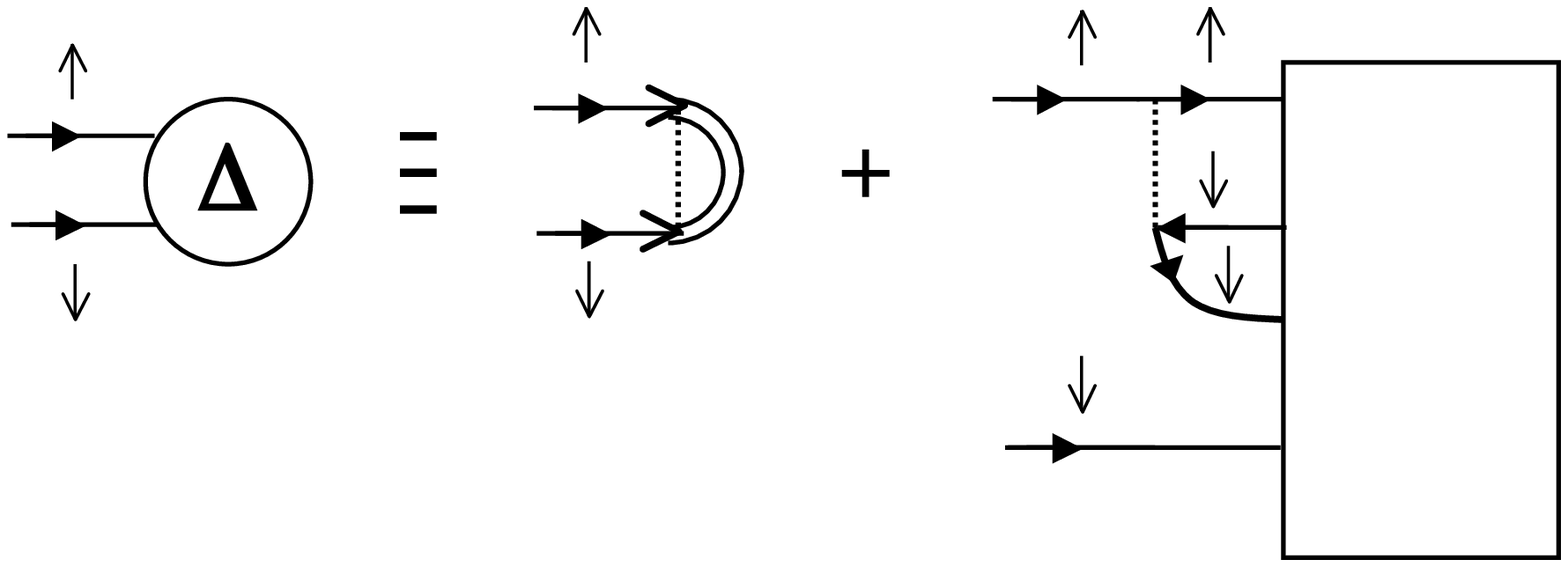}
\caption{The two sets of diagrams making up the anomalous self-energy. By definition the second set excludes
the diagrams which are already included in the first set. In this second set the full propagators drawn may be normal or
anomalous (accordingly we have not drawn the second arrow on these propagators). Similarly the vertex
corresponding to the rectangular box has normal and anomalous parts.}
\label{FigXXA}
\end{figure}

We proceed now in a somewhat similar way for the anomalous self-energy. The resulting equation is the generalization of the standard 'gap equation' of the conventional BCS theory and it has the same character of being finally a self-consistent equation for $\Delta (p)$. In writing an expression for $\Delta (p)$, we can in full generality separate all contributing diagrams in two sets. In the first set, for both of the two entering fermionic lines, the first interaction is between these lines themselves. The second set corresponds to the opposite case, where for at least one of these two entering lines the first interaction is with a line which is not the other entering fermionic line 
(see Fig.\ref{FigXXA}). This separation of the diagrams in two classes is completely analogous to the separation of the diagrams, coming in the standard self-energy \cite{agd}, into an Hartree-Fock like set and another set involving the full vertex.

The set of diagrams where the two fermionic lines interact can be immediately written in terms of $F(p)$,
which is analogous (and actually directly related) to $F^{+}(p)$, except that it has, instead of outgoing lines, an ingoing ($p,\uparrow$) line and an ingoing ($-p, \downarrow$) line. It satisfies a Dyson's equation analogous to Eq.~(\ref{eqdyson},\ref{eqdysona}), namely:
\begin{eqnarray}
\label{eqdysonc}
G(p)=G_0(p)+G(p) \Sigma (p) G_0(p) -F(p) \Delta^{*} (p) G_0(p) \\  \label{eqdysond}
F(p)=F(p) \Sigma (-p) G_0(-p) +G(p) \Delta(p) G_0(-p)
\end{eqnarray}
Introducing the Fourier transform $V({\bf q})$ of the bare instantaneous interaction between $\uparrow$-atom and $\downarrow$-atom, this first contribution to the anomalous self-energy is:
\begin{eqnarray}\label{bcs}
\delta_1({\bf p})=- \,\int \frac{d{\bf k}\,d\omega }{(2\pi )^4}\,V({\bf p}-{\bf k})\,F({\bf k},\omega )
\end{eqnarray}
which has naturally just the form of the standard BCS term.

Let us now switch to the other set of diagrams, for which we proceed to an expansion similar to the one carried out in the preceding subsection. Because of particle conservation this expansion involves now odd powers in $(\Delta ,\Delta ^{*}$). We consider first the term proportional to $\Delta (p') $. The 'normal part' of this diagram contains only one $\uparrow $ line and one $\downarrow$ line (with any
number of interactions), since additional lines are not allowed as discussed above. Moreover, because the interaction is instantaneous, there is no possibility of crossing of interactions when the diagram is drawn in time representation. Finally our lines are not allowed to go backward in time since this would correspond to the creation of a particle-hole pair, which is again not allowed in our case where we have a negative chemical potential at $T=0$ (by contrast, for $\mu >0$, such processes could naturally occur). In summary the only possible diagrams are those of repeated interactions between the $\uparrow $ and the $\downarrow $ atom. This corresponds merely to the propagation of a normal dimer. However these diagrams are already generated by the iteration of the term (\ref{bcs}) when $F(k)$ is written as $G_0(p) \Delta(p) G_0(-p)$ from Eq.(\ref{eqdysond}) at lowest order, and they should not be double-counted. Hence there is no term of order $\Delta $ in this set, and our expansion starts with terms containing $\Delta (p') \Delta^{*} (p'') \Delta (p''')$.

The analysis is now somewhat similar to the one made in the preceding subsection \ref{normself}. To obtain the above third order term in the equation for $\Delta (p)$, we have to write, together with the factor $\Delta (p') \Delta^{*} (p'') \Delta (p''')$, all the 'normal state' diagrams which have, in addition to the two entering lines ($p,\uparrow$) and ($-p,\downarrow)$, another couple of entering lines coming from $\Delta^{*} (p'')$. Moreover, we have two couples of outgoing lines, going into $\Delta (p')$ and $\Delta (p''')$. Just as in the preceding subsection, we have a 'no first interaction' restriction for all these couples of lines, including the ($p,\uparrow$) and ($-p,\downarrow)$ ones. As discussed above, in addition to these ingoing and outgoing ones with their continuations, no other lines can appear in these diagrams since they form loops, which are forbidden since we are at $T=0$ and $\mu <0$. We find that such normal state diagrams have already been essentially encountered in Ref.\cite{bkkcl}, where their sum has been called $\Phi(q_1,q_2;p_2, P)$. Here $q_1$ and $q_2$ are for the entering fermionic lines (corresponding to $p$ and $-p$ in our case), the total four-vector for the ingoing and outgoing lines being $2P$. Finally there is, in the definition of $\Phi$, one entering $2P-q_1-q_2$ dimer, while there are two outgoing dimers $P+p_2$ and $P-p_2$.

Just as in the preceding section, we should take into account that, in the general case, we have the weight $\Delta (p') \Delta^{*} (p'') \Delta (p''')$ which depends on the four-vectors of the atom propagators making up the dimers, while in the definition of $\Phi$ the summation over these variables is already performed, with only their sums $2P-q_1-q_2$, $P+p_2$ and $P-p_2$ being fixed. This problem could nevertheless be circumvented and an expression for the 'normal part' obtained in terms of $\Phi$, in a way  analogous to Appendix \ref{appT3}. However, just as in subsection \ref{normself}, this is not necessary. Indeed from the lowest order term  $\delta_1({\bf p})$ given by Eq.(\ref{bcs}), we see that $\Delta (p)$ does not depend at lowest order on the frequency variable $\nu$. Moreover, since we deal with a bare interaction potential $V({\bf r})$ which has a very short range, its Fourier transform $V({\bf q})$ is a constant, except for very large wave vectors. But, as we will see, these wave vectors will not come into play because all the integrals we deal with converge at a much shorter wave vector. Hence we find that, at lowest order, $\Delta (p)$ turns out to be independent of $p$, just as in standard BCS theory. Now, when we want to calculate the third order term, we may insert in it the expression of $\Delta (p)$ to lowest order, which is just a constant which we denote again as $\Delta $. Accordingly the above problem disappears and the 'normal part' of our diagrams can be expressed directly in terms of $\Phi(q_1,q_2;p_2, P)$.

Nevertheless we have to take into account, just as in subsection \ref{normself}, that some of the diagrams appearing in $\Delta\, |\Delta |^2 \Phi$ are actually not irreducible, and should not be included in our expression for $\Delta (p)$. Hence we have to subtract them out of $\Delta\, |\Delta |^2\Phi$. We denote $\Phi '$ what remains of $\Phi$ after this subtraction. This will be taken care of just below. Gathering our results, we end up with the following (self-consistent) equation:
\begin{eqnarray}\label{eqdelta}
\Delta (p)=- \,\int \frac{d{\bf k}\,d\omega }{(2\pi )^4}\,V({\bf p}-{\bf k})\,F({\bf k},i\omega )
+\frac{1}{2} \,\Delta\, |\Delta |^2 \; \Phi '(p,-p;0, 0)
\end{eqnarray}
The factor 1/2 in front of $\Phi '$ is a general topological factor (it is present even if we do not take $\Delta $ as constant). Indeed exchanging
$\Delta (p')$ and $\Delta (p''')$ corresponds to a mere change of variable and not to a different diagram. On the other hand, by making use of $\Phi '(q_1,q_2;p_2, P)$, any given such diagram would appear twice (even if $\Phi$ is actually unchanged when we exchange the bosonic variables of the two outgoing dimers, which amounts to change $p_2$ into $-p_2$). Hence the factor 1/2 is required to avoid this double counting. We note that clearly $\Delta (p)$ depends indeed on $p$ in general, since this is already explicitely the case in Eq.(\ref{eqdelta}) where  the third order term depends on $p$ through $\Phi '(p,-p;0, 0)$.

\begin{figure}
\centering
\includegraphics[width=100mm]{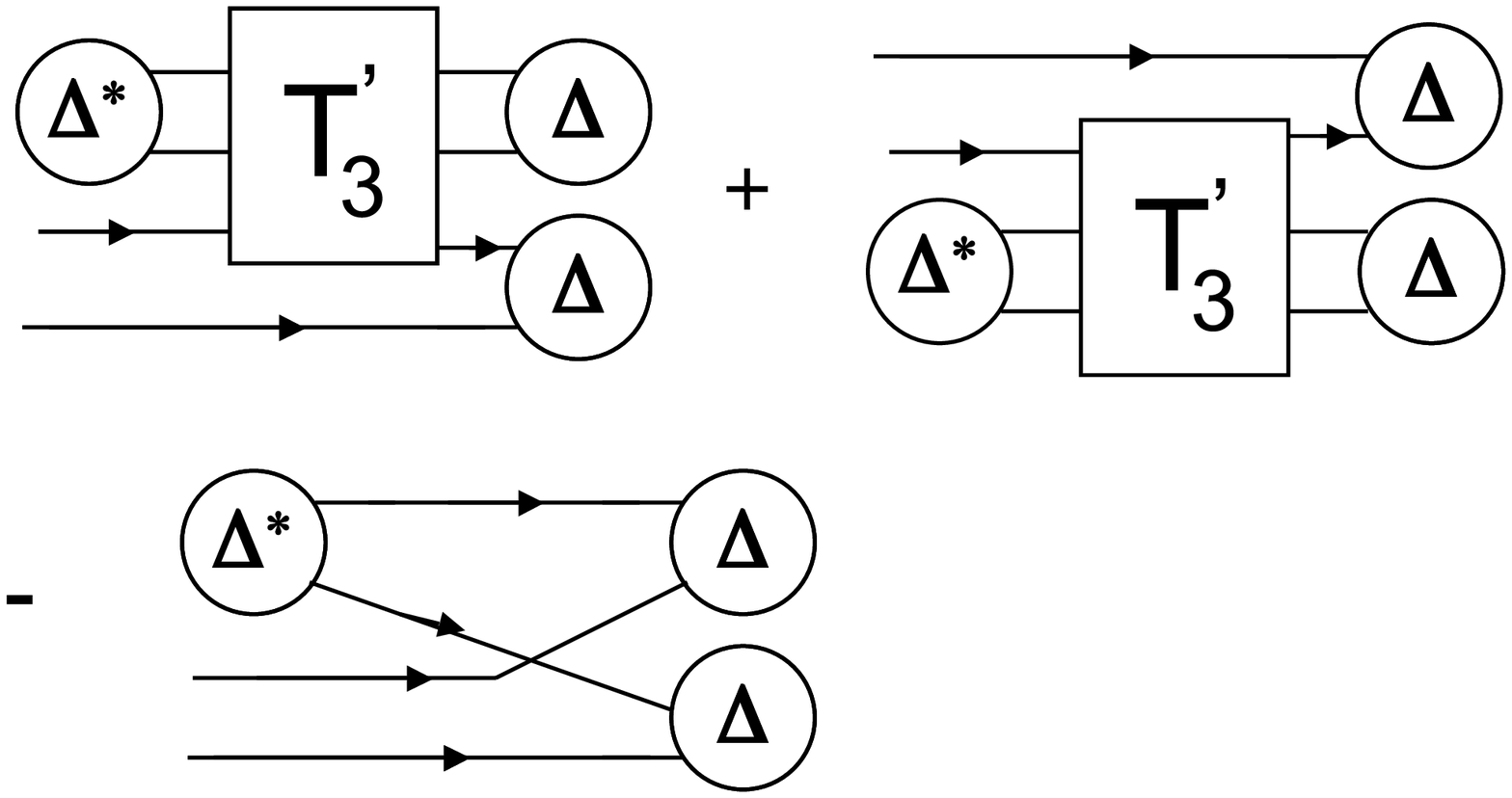}
\caption{Reducible contributions to the anomalous self-energy from $\Delta\, |\Delta |^2 \; \Phi(p,-p;0, 0)$. In the first two terms
$T'_3$ is obtained from $T_3$ by excluding the Born term, since it is taken into account explicitely by the third term.}
\label{FigXXXX}
\end{figure}

Let us now be specific about the difference between $\Phi$ and $\Phi'$, due to the reducible diagrams contained in $\Delta\, |\Delta |^2\Phi$. One can obtain them by looking at the diagrammatic expansion \cite{bkkcl} of $\Phi$. However it is easier to notice, as indicated above, that they appear automatically explicitely from Eq.(\ref{eqdysonc},\ref{eqdysond}), when their solution are expanded in order to obtain the complete perturbative series expansion of $G(p)$ and $F(p)$ in powers of $\Sigma$ and $\Delta $, because this expansion provides precisely all the reducible contributions in terms of the irreducible self-energies $\Sigma$ and $\Delta $. When we expand the solution for $F(p)$:
\begin{eqnarray}
F(p)=\frac{\Delta (p)}{[G^{-1}_0(p)-\Sigma (p)][G^{-1}_0(-p)-\Sigma (-p)]+\Delta (p) \Delta^{*} (p)}
\end{eqnarray}
to third order, we find:
\begin{eqnarray}\label{eqf3}
F(p)=G_0(p)\left[\Delta (p)+\Sigma^{(2)}(p)G_0(p) \Delta (p)+ \Delta (p)G_0(-p)\Sigma^{(2)}(-p)-\Delta (p)G_0(-p) \Delta^{*} (p)G_0(p) \Delta (p)\right]G_0(-p)
\end{eqnarray}
where for consistency we have to take only the second order contribution $\Sigma^{(2)}(p)$ (given by Eq.(\ref{eqsigmn2})) to $\Sigma(p)$. The third order terms in the bracket are clearly the reducible contributions to $\Delta\, |\Delta |^2 \Phi(p,-p;0, 0)$, as it can be seen directly. They are displayed diagrammatically in Fig.\ref{FigXXXX}. In particular the last term of the bracket comes from the Born term of $\Phi(p,-p;0, 0)$. Hence we have explicitely:
\begin{eqnarray}
\frac{1}{2}\Delta\, |\Delta |^2 \Phi '(p,-p;0, 0)&=&\frac{1}{2} \Delta\, |\Delta |^2\Phi (p,-p;0, 0) \\ \nonumber
&&-\left[\Sigma^{(2)}(p)G_0(p) \Delta (p)+ \Delta (p)G_0(-p)\Sigma^{(2)}(-p)-\Delta (p)G_0(-p) \Delta^{*} (p)G_0(p) \Delta (p)\right]
\end{eqnarray}

However it is now more convenient to work with $\Phi$ rather than $\Phi'$. Hence we add to both members of Eq.(\ref{eqdelta}) the reducible contributions. After multiplication by $G_0(p)G_0(-p)$ and taking into account Eq.(\ref{eqf3}), we end up with:
\begin{eqnarray}\label{eqf}
F(p)=G_0(p)G_0(-p)\delta_1({\bf p})
+\frac{1}{2} \,\Delta\, |\Delta |^2 \; G_0(p)G_0(-p) \Phi (p,-p;0, 0)
\end{eqnarray}
Finally we obtain a closed equation by eliminating $F(p)$ in favor of $\Delta (p)$ through Eq.(\ref{eqdelta}), or more simply through the definition Eq.(\ref{bcs}). Multiplying Eq.(\ref{eqf}) by $V({\bf k}-{\bf p})$ and summing over $p$ we find:
\begin{eqnarray}\label{eqgap1}
\delta_1({\bf k})=- \,\int \frac{d{\bf p}\,d\nu }{(2\pi )^4}\,V({\bf k}-{\bf p})\,G_0(p)G_0(-p)\delta_1({\bf p})
-\frac{1}{2} \,\Delta\, |\Delta |^2 \; \,\int \frac{d{\bf p}\,d\nu }{(2\pi )^4}\,V({\bf k}-{\bf p})G_0(p)G_0(-p) \Phi (p,-p;0, 0)
\end{eqnarray}
If we had only the first term in the right-hand side, we would merely have the linearized BCS gap equation.
Accordingly we proceed in the standard way \cite{galit,bel} to eliminate the bare potential in favor of the scattering amplitude. This is more easily understood by proceeding in a formal way. Introducing the operator $\hat{V}$, with matrix elements $V_{{\bf k},{\bf k}'} \equiv V({\bf k}-{\bf k}')$, and the diagonal operator $\hat{D}$ with matrix elements $D_{p,p'} \equiv \delta_{p,p'}G_0(p)G_0(-p)$, the $T$-matrix for two fermions (in the center of mass frame) in vacuum satisfies $\hat{T}=\hat{V}+\hat{V}\hat{D}\hat{T}=\hat{V}+\hat{T}\hat{D}\hat{V}$, where the operator product implies naturally the summation on the intermediate variables with $ \sum_{p}\equiv -(2\pi )^{-4} \int d{\bf p}\,d\nu$. This can just be written $\hat{V}^{-1}=\hat{T}^{-1}+\hat{D}$. If we note $F'(p)$ the second term in the right-hand side of Eq.(\ref{eqf}), we can rewrite Eq.(\ref{eqgap1}) as $\delta_1=\hat{V}\hat{D}\delta_1+\hat{V}F'$, or $\hat{V}^{-1}\delta_1=\hat{D}\delta_1+F'$. Inserting the above value of $\hat{V}^{-1}$, this leads to $\hat{T}^{-1}\delta_1=F'$, or $\delta_1=\hat{T}F'$. This is explicitely:
\begin{eqnarray}\label{eqgap2}
\delta_1({\bf k})= \sum_{p}T({\bf k},{\bf p},E)F'(p)
\end{eqnarray}
where $E$ is the energy at which the $T$-matrix is calculated. In the present case, since $G_0(p)=[i\omega -\epsilon _k+\mu  ]^{-1}$, the total energy for the two fermions is $E=2\mu $ ($\mu $ for each fermion), as it can be seen directly by integrating the first term of Eq.(\ref{eqgap1}) over $\nu$.

In standard scattering theory the scattering amplitude $f_{{\bf k}}({\bf k}')$ at energy $E=k^2/m=k'^2/m$ (the reduced mass is $m/2$) is obtained from the evaluation of the $T$-matrix "on the shell", i.e. taking $T({\bf k},{\bf k}',E)$ with $E=k^2/m$. In the present case we deal with a potential with a short range of order $r_0$. This implies that $k$ and $k'$ have to vary on a typical scale $1/r_0$ in order to have a significant variation of $T({\bf k},{\bf k}',E)$. Since in the case of ultracold atoms we are dealing with much smaller wavevectors, and we may replace $T({\bf k},{\bf k}',E)$ by $T({\bf 0},{\bf 0},E)$. Hence the scattering amplitude is merely given by $f_{{\bf k}}({\bf k}')= -(m/4\pi) \,T({\bf 0},{\bf 0},E)$. On the other  hand, in this same regime, the scattering amplitude is given by its s-wave component $f(k)=-[1/a-\sqrt{-mE}]^{-1}$. This gives $T({\bf 0},{\bf 0},E)=(4\pi/m )[1/a-\sqrt{-mE}]^{-1}$. 

Now we evaluate Eq.(\ref{eqgap2}) for ${\bf k}$ small compared to $1/r_0$. We see that in this range $\delta_1({\bf k})$ is essentially a constant, which we call simply $\delta_1$. On the other hand $F'(p)$ is a rapidly decreasing function \cite{bkkcl} of ${\bf p}$, on a typical scale $1/a$. Hence we may replace in Eq.(\ref{eqgap2}) $T({\bf k},{\bf p},E)$ by $T({\bf 0},{\bf 0},2\mu )$, which can be expressed in terms of the scattering amplitude. This leads explicitely to:
\begin{eqnarray}\label{eqgap3}
\frac{m}{4\pi}\left[\frac{1}{a}-\sqrt{2m|\mu |}\right]\,\delta_1=\frac{1}{2}  \,\Delta\, |\Delta |^2  \sum_{p}G_0(p)G_0(-p) \Phi (p,-p;0, 0)
\end{eqnarray}
Finally we take into account the fact that the sum over $p$ appearing in the right-hand side of Eq.(\ref{eqgap3}) is directly
related to the dimer-dimer scattering amplitude $T_4$ in vacuum \cite{bkkcl} by:
\begin{eqnarray}\label{eqt4}
T_4(0,0;\{{\bf 0},2\mu \})=\sum_{p} G_0(-p)G_0(p)\Phi(-p,p;0,0)
\end{eqnarray}
whereas:
\begin{eqnarray}\label{eqaM}
\left(\frac{8\pi}{m^2a}\right)^2T_4(0,0;\{{\bf 0},-E_b\})=\frac{2\pi(2a_M)}{m}.
\end{eqnarray}
We note that, in Eq.(\ref{eqt4}), we end up with $T_4$ being evaluated at a dimer energy $2\mu $, while the dimer-dimer scattering in
Eq.(\ref{eqaM}) is related to $T_4$ evaluated at the binding energy $-E_b=-1/ma^2$. However at this stage the small
difference between $2\mu $ and $-1/ma^2$ is unimportant in $T_4$, since it appears in the corrective term which is on the right-hand side of Eq.(\ref{eqgap3}). In this term we may also, within the same level of accuracy, use $\delta_1 \simeq \Delta$. This leads finally to:
\begin{eqnarray}\label{mean}
\frac{1}{a}-\sqrt{2m|\mu|}&=&\frac{m^2 a^2}{8}\, a_M\,|\Delta|^2
\end{eqnarray}
When we substitute in this expression the relation $|\Delta |^2 \simeq 8\pi n/m^2 \,a$ found in section \ref{deltexp} A, we obtain at the same level of accuracy:
\begin{eqnarray}\label{eqmean}
\mu = - \frac{1}{2ma^2}\,+ \frac{\pi n\,a_M}{m}
\end{eqnarray}
which is the standard mean field correction to the chemical potential, with the proper dimer-dimer scattering length
$a_M$.

\section{collective mode}\label{collmod}
Since the LHY terms come physically from the low energy collective mode, we need first to derive the expression for the collective mode propagator, as well as its anomalous counterpart. These propagators can be seen as the extension to the superfluid state of the dimer propagator in the normal state. For the normal component of this propagator, this is just the full vertex with two entering and two outgoing fermion lines, the two fermions having opposite spins. In the normal state, in the dilute limit, this vertex is just the single dimer propagator $T_2$ already used in Ref.\cite{bkkcl}. It depends only
on the total momentum $\textbf{P}$ and the total energy $\Omega $ of the two fermions. In contrast with the above reference we work in the grand canonical ensemble and the energies are taken with respect to the chemical potential $\mu $ of single fermions. Hence for all the energies $e $ used in Ref.\cite{bkkcl},
we have to make the substitution $e \rightarrow \omega +\mu $. For the single dimer propagator we have to make $E \rightarrow \Omega +2 \mu $. However we keep for simplicity the same notation as in Ref.\cite{bkkcl} for the various quantities after having made this substitution. Also for coherence, and in contrast with the preceding sections, 
we work in the present section with real frequencies, rather than with Matsubara frequencies.
This is also more natural physically since we are interested in particular in the (real) frequency of the collective mode.

Making use of the quadrivector notation $P = \{{\textbf P},\Omega \}$ we have thus, for this single dimer propagator in the
dilute limit:
\begin{eqnarray}\label{2Vertex}
T_2(P) = \frac{4\pi}{m^{3/2}}\,
\frac{1}{\sqrt{E_b}-\sqrt{{\mathbf P}^2/4m-\Omega + 2 |\mu |}}=
\frac{4\pi}{m^{3/2}}\,
\frac{\sqrt{E_b}+\sqrt{{\mathbf P}^2/4m-\Omega + 2 |\mu |}}{\Omega - 2 |\mu |- {\mathbf P}^2/4m+E_b}
\end{eqnarray}
where $E_b=1/ma^2$ is the dimer binding energy, and we have made explicit the fact that $\mu  <0$. Note that physically this propagator describes the bound state of the two fermions, through its pole at
$\Omega = 2 |\mu | + {\mathbf P}^2/4m-E_b$ , as well as the continuum of scattering states where the dimer is broken, through the continuum in the spectral function starting at $\Omega = {\mathbf P}^2/4m + 2 |\mu |$.

\begin{figure}
\centering
\includegraphics[width=100mm]{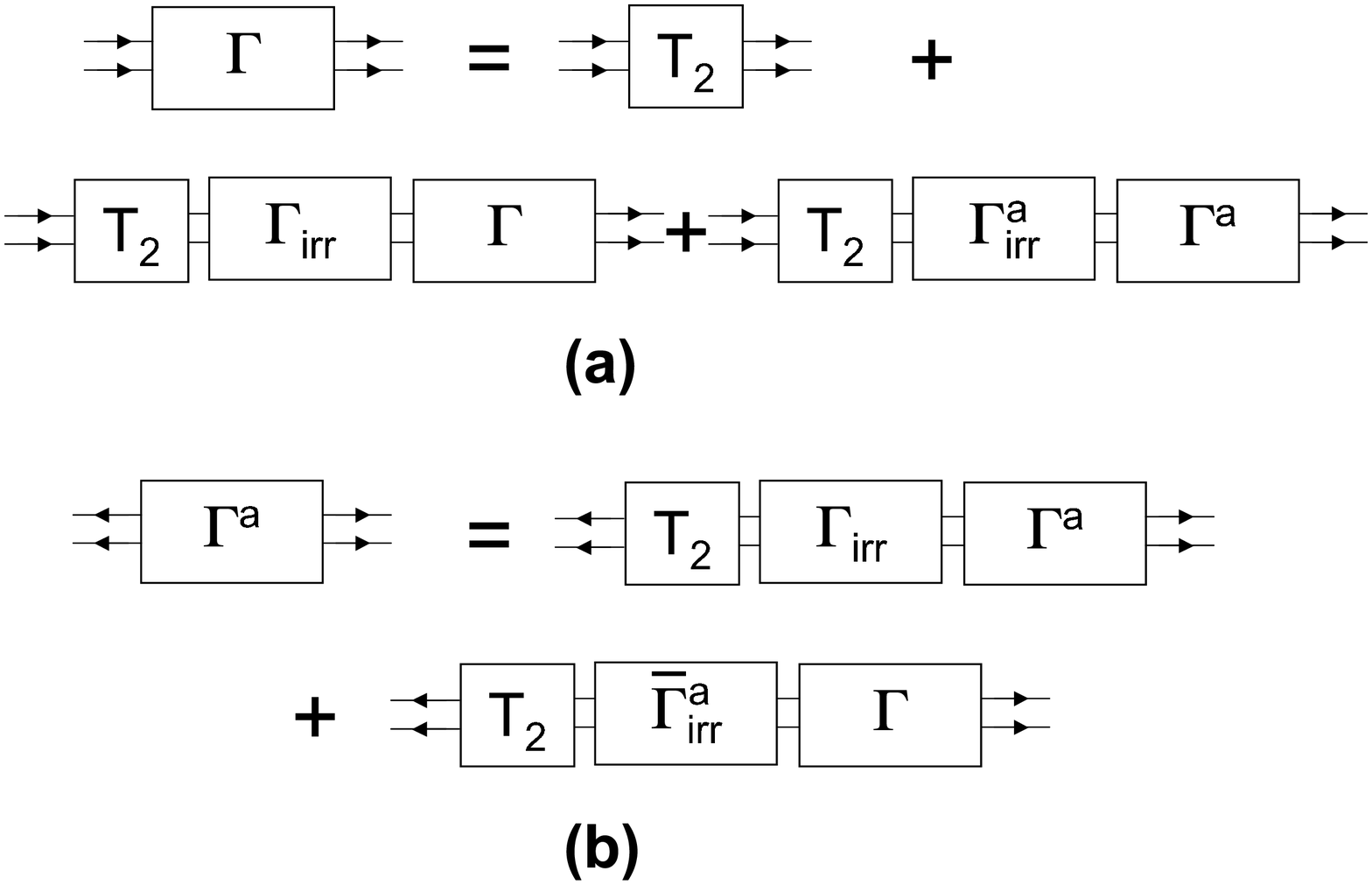}
\caption{Bethe-Salpeter equations for the full vertex in the particle-particle channel.
a) Normal part. b) Anomalous part.}
\label{FigGammaab}
\end{figure}

In the superfluid we want to obtain the correction to this normal state result to lowest order in $|\Delta |^2$. More precisely we want to correct what corresponds to the self-energy if we had a single particle propagator. In the present case this is the irreducible vertex in the particle-particle channel, which by definition can not be separated in two disconnected parts by cutting two fermion propagators. More precisely, since the bare interaction between fermions is already taken into account in $T_2(P)$, this is the part of this irreducible vertex beyond the bare interaction we are interested in. This part $\Gamma_{\mathrm irr}(q_1,q_2;q'_1,q'_2)$ of the irreducible vertex depends on the quadrivectors $q_1$ and $q_2$ of the entering particles, as well as on those $q'_1$ and $q'_2$ of the outgoing particles. Assuming for the moment quite generally that the full vertex 
$\Gamma(p_1,p_2;p'_1,p'_2)$ depends also on these four quadrivectors, we can write a Bethe-Salpeter equation, which is shown diagrammatically in Fig.~\ref{FigGammaab}a:
\begin{eqnarray}
\hspace{-10mm}\Gamma(p_1,p_2;p'_1,p'_2)=T_2(p_1+p_2=p'_1+p'_2) +  \sum_{q_1,q'_1} 
T_2(p_1+p_2=q_1+q_2) \,\Gamma_{\mathrm irr}(q_1,q_2;q'_1,q'_2)\, \Gamma(q'_1,q'_2;p'_1,p'_2)
\nonumber
\end{eqnarray}
\vspace{-5mm}
\begin{eqnarray}\label{betsal}
\hspace{40mm}+  \sum_{q_1,q'_1} 
T_2(p_1+p_2=q_1+q_2) \,\Gamma^a_{\mathrm irr}(q_1,q_2;-q'_1,-q'_2)\, \Gamma^a(-q'_1,-q'_2;p'_1,p'_2)
\end{eqnarray}
where $\sum_{q}\equiv i \int d^3qd\Omega/(2\pi)^4$. We have made explicit the fact that the normal state dimer propagator $T_2(P)$ depends only on the sum $P=p_1+p_2=p'_1+p'_2$ of the incoming and outgoing fermions quadrivectors. In Eq.(\ref{betsal}) the first sum is completely analogous to what should be written in the normal state. However we had also to write the second sum, which describes the anomalous processes where a dimer can be annihilated into the condensate or can be created from the condensate. This leads to introduce, in correspondance with the first sum, an anomalous irreducible vertex $\Gamma^a _{\mathrm irr}$ where four fermions are entering (annihilated into the condensate) and none is going out. Correspondingly we need to introduce an anomalous full vertex $\Gamma^a$ with only outgoing fermion lines. Naturally in order to close our set of equations we have to write for this vertex the corresponding
Bethe-Salpeter equation, which is shown diagrammatically in Fig.~\ref{FigGammaab}b, the major difference with the preceding one being that there is no normal state term:
\begin{eqnarray}
\hspace{-10mm}\Gamma^a(-p_1,-p_2;p'_1,p'_2)=  \sum_{q_1,q'_1} 
T_2(-p_1-p_2=-q_1-q_2) \,\Gamma_{\mathrm irr}(-q'_1,-q'_2;-q_1,-q_2)\, 
\Gamma^a(-q'_1,-q'_2;p'_1,p'_2)
\nonumber
\end{eqnarray}
\vspace{-5mm}
\begin{eqnarray}\label{betsal1}
\hspace{40mm}+  \sum_{q_1,q'_1} 
T_2(-p_1-p_2=-q_1-q_2) \,{\bar \Gamma}^a_{\mathrm irr}(-q_1,-q_2;q'_1,q'_2)\, \Gamma(q'_1,q'_2;p'_1,p'_2)
\end{eqnarray}
where we have introduced the irreducible vertex ${\bar \Gamma}^a_{\mathrm irr}$ where there are only outgoing fermions, physically created from the condensate.

Now it is explicit in Eq.(\ref{betsal}) that  $\Gamma(p_1,p_2;p'_1,p'_2)$ depends only on $P=p_1+p_2$. By time reversal symmetry it depends also only on $p'_1+p'_2=P$, this last equality resulting from momentum-energy conservation. Hence, just as $T_2(P)$, it depends only on the total momentum and energy $P$, and we merely denote this function as $\Gamma (P)$. Similarly $\Gamma^a(-p_1,-p_2;p'_1,p'_2)$ depends only on $-P=-p_1-p_2=p'_1+p'_2$ and we call it merely $\Gamma^a (-P)$. In this way Eq.(\ref{betsal}) becomes simply:
\begin{eqnarray}\label{betsalp}
\Gamma(P)=T_2(P) + 
T_2(P) \,\Gamma_{\mathrm irr}(P)\, \Gamma(P)
+T_2(P) \,\Gamma^a_{\mathrm irr}(P)\, \Gamma^a(-P)
\end{eqnarray}
and similarly:
\begin{eqnarray}\label{betsalp1}
\Gamma^a(-P)=
T_2(-P) \,\Gamma_{\mathrm irr}(-P)\, \Gamma^a(-P)
+T_2(-P) \,{\bar \Gamma}^a_{\mathrm irr}(-P)\, \Gamma(P)
\end{eqnarray}
where we have introduced:
\begin{eqnarray}\label{gamir}
\Gamma_{\mathrm irr}(P) \equiv \sum_{q_1,q'_1} \Gamma_{\mathrm irr}(q_1,P-q_1;q'_1,P-q'_1) \\
\Gamma^a_{\mathrm irr}(P) \equiv \sum_{q_1,q'_1} \Gamma^a_{\mathrm irr}(q_1,P-q_1;-q'_1,-P+q'_1) \\
{\bar \Gamma}^a_{\mathrm irr}(P) \equiv \sum_{q_1,q'_1} {\bar \Gamma}^a_{\mathrm irr}(-q_1,-P+q_1;q'_1,P-q'_1)
\end{eqnarray}
These equations are naturally quite similar to the ones we had for the single fermion Green's function. Their solutions are:
\begin{eqnarray}\label{gamn}
\Gamma(P)=
\frac{T_2^{-1}(-P)-\Gamma_{\mathrm irr}(-P)}
{\left[T_2^{-1}(P)-\Gamma_{\mathrm irr}(P)\right]
\left[T_2^{-1}(-P)-\Gamma_{\mathrm irr}(-P)\right]
-\Gamma^a_{\mathrm irr}(P){\bar \Gamma}^a_{\mathrm irr}(-P)}
\end{eqnarray}
and
\begin{eqnarray}\label{gama}
\Gamma^a(-P)=
\frac{{\bar \Gamma}^a_{\mathrm irr}(-P)}
{\left[T_2^{-1}(P)-\Gamma_{\mathrm irr}(P)\right]
\left[T_2^{-1}(-P)-\Gamma_{\mathrm irr}(-P)\right]
-\Gamma^a_{\mathrm irr}(P){\bar \Gamma}^a_{\mathrm irr}(-P)}
\end{eqnarray}
As it could be expected physically, these equations are formally identical to those obtained for pure bosons \cite{agd}.
In the same spirit we note, for later use, that we could also introduce an anomalous full vertex $\bar{\Gamma}^a$ with only ingoing fermion lines. It is easily shown, by writing the equivalent of Eq.(\ref{betsalp}) with this vertex, to be related to the preceding ones by $\bar{\Gamma}^a(P)\,{\bar \Gamma}^a_{\mathrm irr}(-P)=\Gamma^a(-P)\,\Gamma^a_{\mathrm irr}(P)$.
This gives:
\begin{eqnarray}\label{gamabar}
\bar{\Gamma}^a(P)=
\frac{\Gamma^a_{\mathrm irr}(P)}
{\left[T_2^{-1}(P)-\Gamma_{\mathrm irr}(P)\right]
\left[T_2^{-1}(-P)-\Gamma_{\mathrm irr}(-P)\right]
-\Gamma^a_{\mathrm irr}(P){\bar \Gamma}^a_{\mathrm irr}(-P)}
\end{eqnarray}

We have now to find the expressions of $\Gamma_{\mathrm irr}$, $\Gamma^a_{\mathrm irr}$ and ${\bar \Gamma}^a_{\mathrm irr}$
within our expansion in powers of $\Delta $. Since $\Delta $ corresponds to diagrams with two ingoing fermions and $\Delta ^{*}$ to two outgoing fermions, particle conservation implies that the expansions start at second order with $\Gamma_{\mathrm irr}$ proportional to  $\Delta \Delta ^{*}$, $\Gamma^a_{\mathrm irr}$ proportional to  $\Delta ^2$ and ${\bar \Gamma}^a_{\mathrm irr}$ proportional to  $\Delta ^{*2}$. Since it can be seen that it is not required to go to higher order, the expressions of our irreducible vertices will be given by these second order expressions. Since we have written explicitely the various factors $\Delta $ and $\Delta^{*}$, the coefficients in the expressions of $\Gamma_{\mathrm irr}$,$\Gamma^a_{\mathrm irr}$ and ${\bar \Gamma}^a_{\mathrm irr}$ correspond to diagrams containing only normal state propagators. 

\begin{figure}
\centering
\includegraphics[width=120mm]{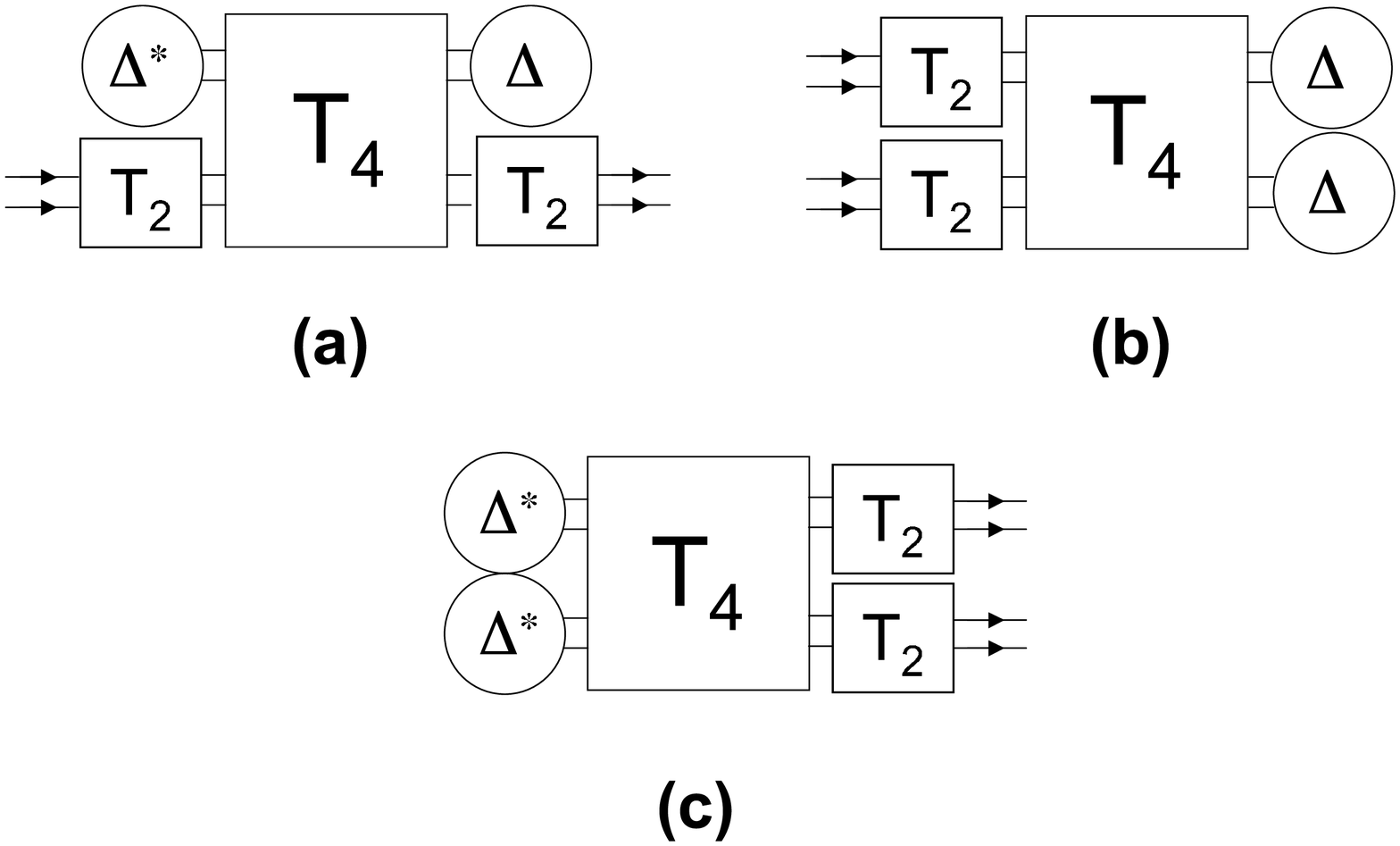}
\caption{Diagrammatic representation for the irreducible vertices coming in the collective mode propagator, at second order
in $\Delta $.
a) Normal irreducible vertex $\Gamma_{\mathrm irr}(P)$. b) and c) Anomalous irreducible vertices
$\Gamma^a_{\mathrm irr}(P)$ and ${\bar \Gamma}^a_{\mathrm irr}(P)$.}
\label{FigXXabc}
\end{figure}

For $\Gamma_{\mathrm irr}$ (see Fig.~\ref{FigXXabc}a), we have to write the expression for all the normal state diagrams with one ingoing dimer propagator and one outgoing dimer (from the very definition of $\Gamma_{\mathrm irr}$). Moreover there are in addition two lines ingoing from $\Delta ^{*}$ and two lines outgoing to $\Delta $. These last two couples of lines have to be treated as dimer lines, which means that they are not allowed to interact immediately in the normal state diagrams we are drawing, because such interactions are already taken into account in $\Delta $ and $\Delta ^*$, and allowing them would amount to double counting. In summary we have to write all possible normal state diagrams with two ingoing dimer lines and two outgoing dimer lines (these diagrams are automatically irreducible in the sense of the definition of $\Gamma_{\mathrm irr}$, that is with respect to a single dimer propagator, since the four ingoing normal state lines can not disappear). This is depicted diagrammatically in Fig.~\ref{FigXXabc}a. But this set of diagrams is just by definition the normal state dimer-dimer scattering vertex $T_4$ introduced in \cite{bkkcl}. More precisely, since the total momentum-energy linked to $\Delta $ and $\Delta ^*$ is zero and the total momentum-energy of the ingoing lines is $P$, this vertex is $T_4(P/2,P/2;P/2)$ with the notations of Ref.\cite{bkkcl} which gives for $\Gamma_{\mathrm irr}$ the explicit expression:
\begin{eqnarray}\label{gammirr}
\Gamma_{\mathrm irr}(P)=|\Delta |^2 \;T_4\left(\frac{P}{2},\frac{P}{2};\frac{P}{2}\right)
\end{eqnarray}
With completely analogous arguments for $\Gamma^a_{\mathrm irr}$, we obtain:
\begin{eqnarray}\label{gammirra}
\Gamma^a_{\mathrm irr}(P)=\frac{1}{2}\,\Delta^2 \;T_4\left(P,0;0\right)
\end{eqnarray}
and similarly:
\begin{eqnarray}\label{gammirrba}
{\bar \Gamma}^a_{\mathrm irr}(P)=\frac{1}{2}\,\Delta^{*2} \;T_4\left(0,P;0\right)
\end{eqnarray}
corresponding to the diagrams found in Fig.~\ref{FigXXabc}b and Fig.~\ref{FigXXabc}c. In Eq.(\ref{gammirra}) and (\ref{gammirrba}) we had to put a 1/2 topological
factor corresponding to the fact that, for example, exchanging the two factors $\Delta $ in Eq.(\ref{gammirra}) (that is precisely the set of diagrams they correspond to) does not produce a different diagram, whereas this diagram would appear twice if we had just written $\Delta^2 \;T_4$.

Finally it is worth pointing out that $T_4$ satisfies $T_4(-p_1,p_2;P)=T_4(p_1,p_2;P)$. This is physically obvious since it corresponds to state that exchanging the arguments of the two ingoing dimer lines does not change the value of the vertex, since we
exchange two bosons. But this property can also be proved directly from the integral equation \cite{bkkcl} satisfied by $T_4$. Similarly we have $T_4(p_1,-p_2;P)=T_4(p_1,p_2;P)$.

Let us consider now more specifically the collective mode dispersion relation. It is given by the pole of the collective mode propagator, that is:
\begin{eqnarray}\label{moddisp}
{\left[T_2^{-1}(P)-\Gamma_{\mathrm irr}(P)\right]
\left[T_2^{-1}(-P)-\Gamma_{\mathrm irr}(-P)\right]
-\Gamma^a_{\mathrm irr}(P){\bar \Gamma}^a_{\mathrm irr}(-P)}=0
\end{eqnarray}
Naturally we have first to check that for zero wave vector ${\bf P}={\bf 0}$, the frequency $\Omega$ is zero, that is $P=0$ is solution of this dispersion relation. For this case we have from Eq. (\ref{2Vertex}) for the bare dimer propagator:
\begin{eqnarray}\label{2Vertex0}
T_2^{-1}(0) = \frac{m^{3/2}}{4\pi}\,
\left(\sqrt{E_b}-\sqrt{2 |\mu |}\right)
\end{eqnarray}
On the other hand we have 
\begin{eqnarray}\label{gammirr0}
\frac{\Gamma_{\mathrm irr}(0)}{|\Delta |^2}=
\frac{2\Gamma^a_{\mathrm irr}(0)}{\Delta^2}= 
\frac{2{\bar \Gamma}^a_{\mathrm irr}(0)}{\Delta^{*2}} = T_4\left(0,0;0\right)=\frac{1}{16\pi }m^3 a^2\, a_M
\end{eqnarray}
where the last equality comes from Ref.\cite{bkkcl}, and we have used the fact that, in $T_4\left(0,0;\{{\bf 0},-E_b+2|\mu|\}\right)$, we may, at the order we are working, replace $|\mu|$ by its zeroth order value, namely $|\mu|=E_b/2$ (we recall that we have to shift all the frequencies of Ref.\cite{bkkcl} by $2\mu $). Since we have already found that $\sqrt{2 |\mu |}=\sqrt{E_b}-|\Delta |^2 m^{3/2} a^2\, a_M/8$, we can easily check that Eq.(\ref{moddisp}) is indeed satisfied in this case.

Now, as we will see, we are more specifically interested in the collective mode dispersion relation for wave vectors $|{\bf P}| \ll 1/a$ and frequencies $\Omega \ll E_b$. Since $E_b$ and $2|\mu |$ are of the same order of magnitude, we can expand the bare dimer propagator to first order as:
\begin{eqnarray}\label{2Vertex1}
T_2^{-1}(P) = T_2^{-1}(0)-\frac{m^{3/2}}{8\pi \sqrt{2 |\mu |}}\left(\frac{{\bf P}^2}{4m}-\Omega\right)
\end{eqnarray}
On the other hand the typical scale of variation of $T_4$ on its arguments is $1/a$ for the wave vectors and $E_b$ for the energies, since we deal with a normal state quantity and there are accordingly no other scales in the corresponding problem. Hence, for the wave vectors and frequencies we are considering, we can still take:
\begin{eqnarray}\label{gammirr0}
\frac{\Gamma_{\mathrm irr}(P)}{|\Delta |^2} \simeq
\frac{2\Gamma^a_{\mathrm irr}(P)}{\Delta^2} \simeq 
\frac{2{\bar \Gamma}^a_{\mathrm irr}(P)}{\Delta^{*2}} \simeq T_4\left(0,0;0\right)=\frac{1}{16\pi }m^3 a^2\, a_M
\end{eqnarray}
On the other hand the above result at $P=0$ implies also that we have $T_2^{-1}(0)-\Gamma_{\mathrm irr}(0)= -(1/2)\Gamma_{\mathrm irr}(0)=-|\Gamma^a_{\mathrm irr}(0) |$, and Eq. (\ref{moddisp}) for the collective mode frequency
$\Omega_{cm}({\bf P})$ can be rewritten as:
\begin{eqnarray}\label{bog}
\Omega ^2_{cm}({\bf P}) = \left(\frac{{\bf P}^2}{4m}\right)^2 + 2\mu _B \,\frac{{\bf P}^2}{4m}= \left(\frac{{\bf P}^2}{4m}\right)^2 + \frac{\pi n a_M}{m^2} \,{\bf P}^2
\end{eqnarray}
where we have set $\mu _B=4\pi \sqrt{2 |\mu |}\Gamma_{\mathrm irr}(0)/m^{3/2}$ and in the last step we have used the lowest order expressions for $|\mu |$ and $|\Delta |^2$ to calculate this quantity. Naturally Eq.(\ref{bog}) is just the Bogoliubov dispersion relation, with the proper scattering dimer-dimer scattering length.

\section{collective mode contributions to the self-energies}\label{sigcollmod}

As we have already indicated in the introduction, the LHY term arises from the collective mode contributions
to the self-energies. If we were to continue naturally our above procedure and go to next order
in $\Delta $, i.e. to order $\Delta ^4$ for example in the calculation of the normal self-energy, we would find
singular results (as it can be checked by proceeding to such a formal expansion in the explicit calculations
performed below). This is due to the fact that the collective mode is gapless, and the lack of gap in the excitation
spectrum leads to a singularity in the perturbative expansion. Hence, instead of the next order terms, we include
the whole contribution coming from the collective mode. Avoiding in this way to proceed immediately to a $\Delta $ 
expansion is equivalent to a series resummation, which allows to take care of singularities.

Writing the collective mode contributions to the self-energy is formally easy, when one notices for example
the analogy of the normal part of the collective mode
propagator with the product of $\Delta ^{*}$ and $\Delta $. Indeed the term $\Delta ^{*}$ acts as a 'source' of fermions in our diagrams,
coming physically out of the condensate,
while $\Delta $ acts as a 'sink' of fermions going into the condensate. These source and sink related to the condensate
are required since, at $T=0,\mu <0$, no fermions are present except coming from the superfluid. However when we
allow to insert the normal part of a collective mode propagator, the starting part of the propagator acts as a sink 
for the rest of the diagram while the end part
acts as a source. Hence we have just to substitute $\Delta \Delta ^{*}$ by the normal part of
a collective mode propagator. We will proceed in a similar way for the anomalous parts. Naturally we will have
to take care that the self-energy contributions already calculated above are not counted twice, by removing the corresponding
terms. 

\subsection{Contribution to the normal self-energy}\label{modnormal}
\begin{figure}
\centering
\includegraphics[width=60mm]{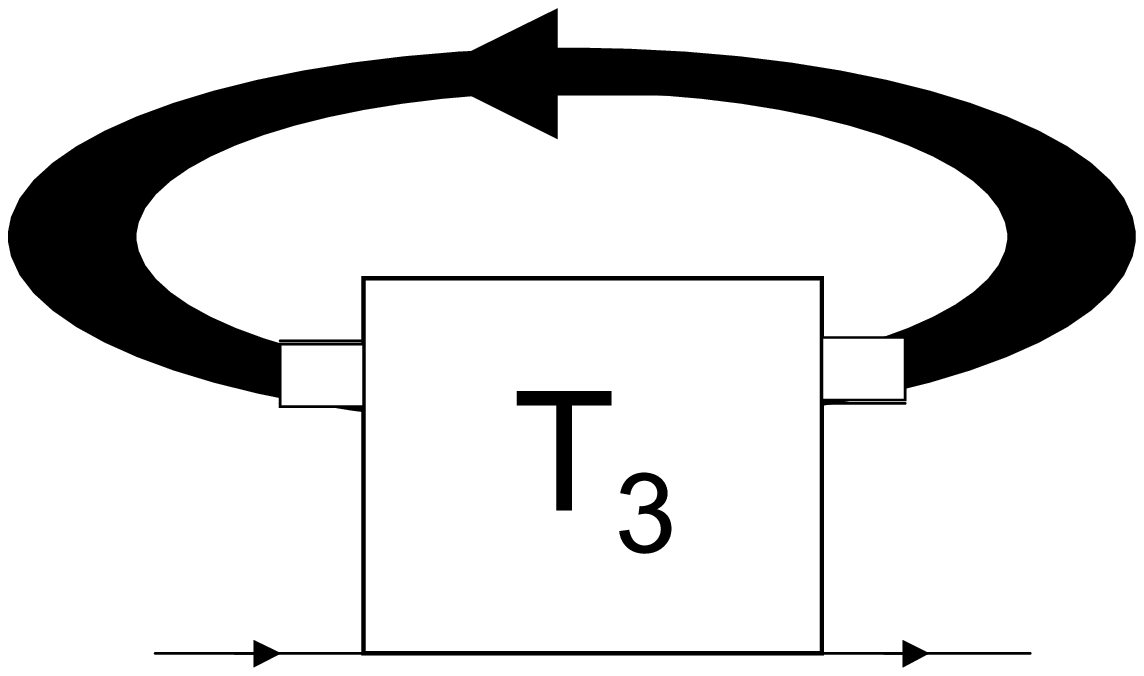}
\caption{Diagrammatic representation of $\Sigma_{cm}(p)$. The diagonal component $\Gamma(P)$
of the collective mode propagator is depicted by the heavy line}
\label{FigXAB}
\end{figure}

Hence the collective mode contribution $\Sigma_{cm}(p)$ to the normal self-energy is obtained by replacing,
in our lowest order expression, the product $\Delta \Delta ^{*}$ by the collective mode propagator, more specifically its 
diagonal component $\Gamma(P)$ given by Eq.(\ref{gamn}), 
as depicted diagrammatically in Fig.\ref{FigXAB}. We have naturally to sum over the variable $P$ of the collective mode. 
Corresponding to Eq.(\ref{eqsigmn2}), this leads to:
\begin{eqnarray}\label{eqsigmod}
\Sigma_{cm}(p)= \sum_{P}\,T_3(p,p;p+P) \Gamma(P)
\end{eqnarray}

In principle this expression is improper because it contains, when we perform its expansion in powers of $|\Delta |^2$, contributions of zeroth and first order in  $|\Delta |^2$ which have already been taken explicitely into account in our above expansion. However in the present case these two contributions are zero. For the zeroth order term, this is fairly obvious physically since it corresponds to a normal state diagram with a normal state dimer propagator, and at $T=0$ no dimers are present which makes all diagrams of this kind equal to zero. Mathematically this would happen because, in this diagram, the presence of a normal state dimer would necessarily imply, since there is no down-spin entering line, the presence of a down-spin loop (if we calculate the up-spin self-energy) which can not be present as explained in Appendix \ref{appnoloop}.

\begin{figure}
\centering
\includegraphics[width=60mm]{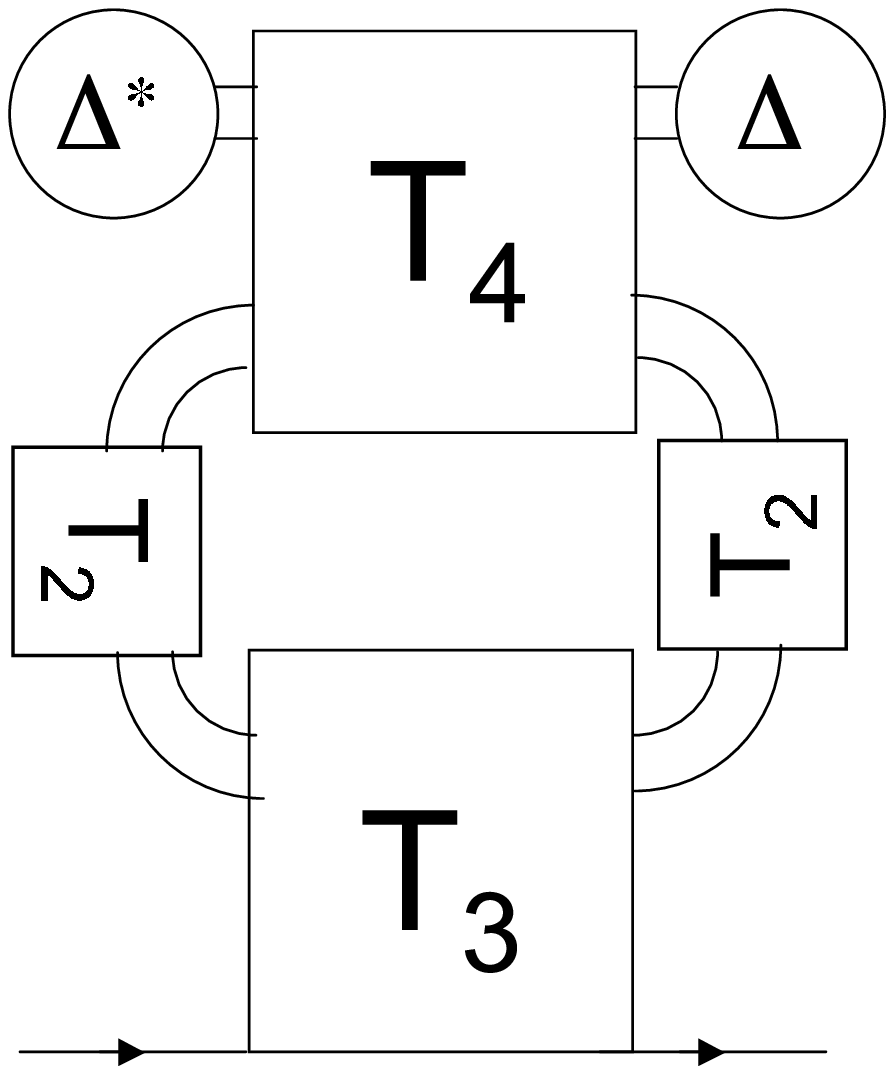}
\caption{First order term in $|\Delta |^2$ in the collective mode contribution to the normal self-energy Eq.(\ref{eqsigmod})
depicted in Fig.\ref{FigXAB}}
\label{FigAAAA}
\end{figure}

The conclusion is the same for the first order term, but the argument is slightly more complicated. The factor of $|\Delta |^2$ in this term is  $ \sum_{P}T_3(p,p;p+P)[T_2(P)]^2T_4(P/2,P/2;P/2)$, which is a normal state diagram (see Fig.\ref{FigAAAA}). It is actually a part of the diagrams belonging to $T_3(p,p;p)$ since it has one entering fermion line and one entering dimer line, with the same for the outgoing lines. However we have
seen, as it is detailed in Appendix \ref{appT3anal}, that only the Born term in $T_3$ gives a non zero contribution to the particle number. On the other hand the term $ \sum_{P}T_3(p,p;p+P)[T_2(P)]^2T_4(P/2,P/2;P/2)$ does obviously not contain the Born term of $T_3$ since this last one reduces to a simple propagator, whereas the presence of the factor $T_2(P)$ implies at least two lines, with interaction between them. Actually one can see more precisely that our term is equal to zero, since it contains necessarily propagators going backward in time (see Appendix \ref{appnoloop} and \ref{appT3anal}). Taking the Born contribution for $T_3$ and $T_4$ in $ \sum_{P}T_3(p,p;p+P)[T_2(P)]^2T_4(P/2,P/2;P/2)$ gives a good example of this. Finally, after these considerations, we see that we can still use Eq.(\ref{eqsigmod}) for 
$\Sigma_{cm}(k)$. The conclusion will be different for the mode contribution to the anomalous self-energy discussed below.

Now when we look for the contribution $n_{cm}$ to the density corresponding to Eq.(\ref{eqsigmod}), we have  to replace $G(p)$ in
Eq.(\ref{eqdefn}) by $G_0(p)\Sigma_{cm}(p)G(p)$,
just as it is done below Eq.(\ref{eqdefn}). Quite similarly it will be enough to replace here $G(p)$ by $G_0(p)$ to obtain the lowest order contribution of the collective mode we are interested in. This gives:
\begin{eqnarray}\label{eqnmod}
n_{cm}= \sum_{p}\Sigma_{cm}(p)\left[G_0(p)\right]^2=
 \sum_{p,P}\,T_3(p,p;p+P) \,\Gamma(P)\left[G_0(p)\right]^2
\end{eqnarray}
where again $ \sum_{p}\equiv -(2\pi )^{-4} \int d{\bf p}\,d\nu $ and $p \equiv ({\bf p},i\nu )$. Now the argument runs just as below Eq.(\ref{eqdefn1}). We integrate first on the frequency variable $\nu $ of the quadrivector $p$. As shown in Appendix \ref{appT3anal}, it turns out that, except for the Born contribution, $T_3(p,p;p+P)$ is analytical in the upper complex half-plane of the variable $\nu $. Since 
$\left[G_0(p)\right]^2=[i\nu -\xi_{\bf p} ]^{-2}$ is also analytical in this half-plane, one obtains by closing the
integration contour at infinity in this half-plane that only the Born part $-G_0(P-p)$ of $T_3(p,p;p+P)$ contributes to the result. Hence for the integration over $\nu $, we have just to calculate:
\begin{eqnarray}
\int_{-\infty}^{\infty}\frac{d\nu }{2\pi }\,\frac{1}{[i\nu -|\mu |-\epsilon _{\bf p} ]^{2}}\,\frac{1}{i\Omega-i\nu- |\mu |-\epsilon _{{\bf P}-{\bf p}}}=
-\frac{1}{[i\Omega -2|\mu |-\epsilon _{\bf p}-\epsilon _{{\bf P}-{\bf p}} ]^{2}}
\end{eqnarray}
and we are left with:
\begin{eqnarray}\label{eqnmoda}
n_{cm}= - \int \frac{d{\bf p}d{\bf P}}{(2\pi )^6} \int_{-\infty}^{\infty}\frac{d\Omega }{2\pi }
\,\frac{\Gamma(P)}{[i\Omega -2|\mu |-\epsilon _{\bf p}-\epsilon _{{\bf P}-{\bf p}} ]^{2}}
\end{eqnarray}
The factor of $\Gamma(P)$ is analytical in the upper complex half-plane of $\Omega$ (which corresponds to the negative frequency half-plane when we go back to the standard frequency language). Hence by closing the $\Omega $ contour in this half-plane, we are left to consider the singularities of $\Gamma(P)$. Physically they correspond to the two-fermions excitations of the system, that is the collective mode and the broken molecule excited states. In the standard frequency language, the corresponding singularities are on the real frequency axis (the imaginary axis for $\Omega $), on the negative as well as on the positive side.

Let us first consider the low frequency singularities. For a fixed value of ${\bf P}$, we have just as singularity a single pole corresponding to the collective mode frequency. Let us call $\Omega_{cm}({\bf P})$ the positive solution of Eq.(\ref{bog}). Our pole is located at $\Omega = i \Omega_{cm}({\bf P})$. In Eq.(\ref{gamn}), which gives
$\Gamma(P)$, for low frequency and wavevector, the denominator becomes $m^3(\Omega^2_{cm}({\bf P})-\Omega^2)/(128\pi ^2|\mu |)$, while from Eq.(\ref{2Vertex1}) the numerator is $-[(1/2)\Gamma_{\mathrm irr}(0)+m^{3/2}({\bf P}^2/4m+\Omega)/(8\pi \sqrt{2 |\mu |})]$. After going to imaginary frequency $\Omega \rightarrow i \Omega$, this gives in $\Gamma(P)$, for the pole at $i \Omega_{cm}({\bf P})$, a residue:
\begin{eqnarray}
Z({\bf P})=\frac{4\pi \sqrt{2 |\mu |}}{i\,m^{3/2}}
\left[1-\frac{\frac{{\bf P}^2}{4m}+\mu _B}{\Omega_{cm}({\bf P})}\right]
\end{eqnarray}
On the other hand, for small $\Omega $ and ${\bf P}$, the ${\bf p}$ integration in Eq.(\ref{eqnmoda}) reduces to:
\begin{eqnarray}\label{intpk}
\int \frac{d{\bf p}}{(2\pi )^3}
\,\frac{1}{[2|\mu |+2\epsilon _{\bf p}]^{2}}=\frac{m^{3/2}}{8\pi \sqrt{2 |\mu |}}
\end{eqnarray}

This leads to:
\begin{eqnarray}\label{eqintx}
n_{cm}= -\frac{m^{3/2}}{8\pi \sqrt{2 |\mu |}} \int \frac{d{\bf P}}{(2\pi )^3} \,iZ({\bf P})=\frac{(4m\mu _B)^{3/2}}{4\pi ^2} \int_{0}^{\infty}dx\,x^2\left[\frac{x^2+1}{\sqrt{x^4+2x^2}}-1\right]
\end{eqnarray}
where in the last step we have made the change of variables $P=(4m\mu _B)^{1/2}x$. The numerical integral is $\sqrt{2}/3$ and our final result is:
\begin{eqnarray}\label{ncmfin}
n_{cm}=\frac{(2m\mu _B)^{3/2}}{3\pi ^2} \simeq \frac{(4\pi na_M)^{3/2}}{3\pi ^2}
\end{eqnarray}
where in the last equality we have made use of the lowest order result for $|\Delta |^2$, given by Eq.(\ref{eqdefn3}). Naturally, once we have taken care of $T_3$, we are essentially back to the elementary boson theory
\cite{agd,ps} and one recognizes in Eq.(\ref{ncmfin}) the standard "depletion of the condensate" for the dilute Bose gas, with $\xi=(4m\mu _B)^{-1/2}$ being its coherence length.

We have first considered the contribution of the low frequency singularities. We have now to finish up by showing that the higher frequency ones give a negligible contribution. First we see that, in Eq.(\ref{eqnmoda}), the denominator plays in this respect an unimportant role, since its modulus is larger than $(2|\mu |+\epsilon _{{\bf p}})^2$. Hence to have an upper bound for the contributions we are investigating, we can replace this denominator by this lower bound, which gives after integration over ${\bf p}$ an unimportant factor $m^{3/2}/(4\pi |\mu |^{1/2})$. Hence we are again left with the evaluation of $ \int \!d\Omega \,\Gamma(P)$.

It is now more convenient to go back to real frequencies (equivalent to the change of variables $i\Omega=\omega$). The integration is now over the imaginary $\omega $ axis and is closed in the negative $\omega $ half-plane. The contour can be deformed to merely enclose the negative $\omega $ axis, and corresponding to the singularities on this axis, ${\mathrm Im}\Gamma(P)$ has a jump across this axis. Hence we have:
\begin{eqnarray}\label{eqcorr1}
-\int_{-\infty}^{\infty}\frac{d\Omega }{2\pi }
\,\Gamma(P)=-\int_{0}^{-\infty}\frac{d\omega }{\pi }\,{\mathrm Im}\Gamma(\{{\bf P},\omega+i\epsilon  \})=
\int_{0}^{-\infty}\frac{d\omega }{\pi }
\,\frac{{\mathrm Im}\Gamma^{-1}(\{{\bf P},\omega+i\epsilon  \})}
{|\Gamma^{-1}(\{{\bf P},\omega+i\epsilon  \})|^2}
\end{eqnarray}
Now, from Eq.(\ref{gamn}), $\Gamma^{-1}(\{{\bf P},\omega+i\epsilon  \})$ contains a first term $T_2^{-1}(P)-\Gamma_{\mathrm irr}(P)$ which has all its singularities on the real positive $\omega $ axis, corresponding to the continuous spectrum of the broken dimer (for $T_2(P)$) and of the broken pair of dimers (for $\Gamma_{\mathrm irr}(P)=|\Delta |^2 \;T_4\left(P/2,P/2;P/2\right)$). Hence it is real on the negative $\omega $ axis and does not contribute to ${\mathrm Im}\Gamma^{-1}$ in Eq.(\ref{eqcorr1}).
From Eq.(\ref{gamn}) the second term of $\Gamma^{-1}$
is $-\Gamma^a_{\mathrm irr}(P){\bar \Gamma}^a_{\mathrm irr}(-P)/(T_2^{-1}(-P)-\Gamma_{\mathrm irr}(-P))$, which is directly proportional to $|\Delta |^4$ from the product $\Gamma^a_{\mathrm irr}(P){\bar \Gamma}^a_{\mathrm irr}(-P)$. Accordingly we have proved basically our point since our result, being proportional to $|\Delta |^4 \sim n^2$, is indeed negligible compared to the result Eq.(\ref{ncmfin}), which is proportional to $n^{3/2}$.

Actually, to complete our proof, we have to be more careful and evaluate the coefficient of $|\Delta |^4$. This is quite obvious since Eq.(\ref{ncmfin}) is in fact a contribution contained in Eq.(\ref{eqcorr1}) (we have made no approximation to reach Eq.(\ref{eqcorr1})). This detailed analysis is performed in Appendix \ref{normselfa}.

\subsection{Contribution to the anomalous self-energy}

\begin{figure}
\centering
\includegraphics[width=100mm]{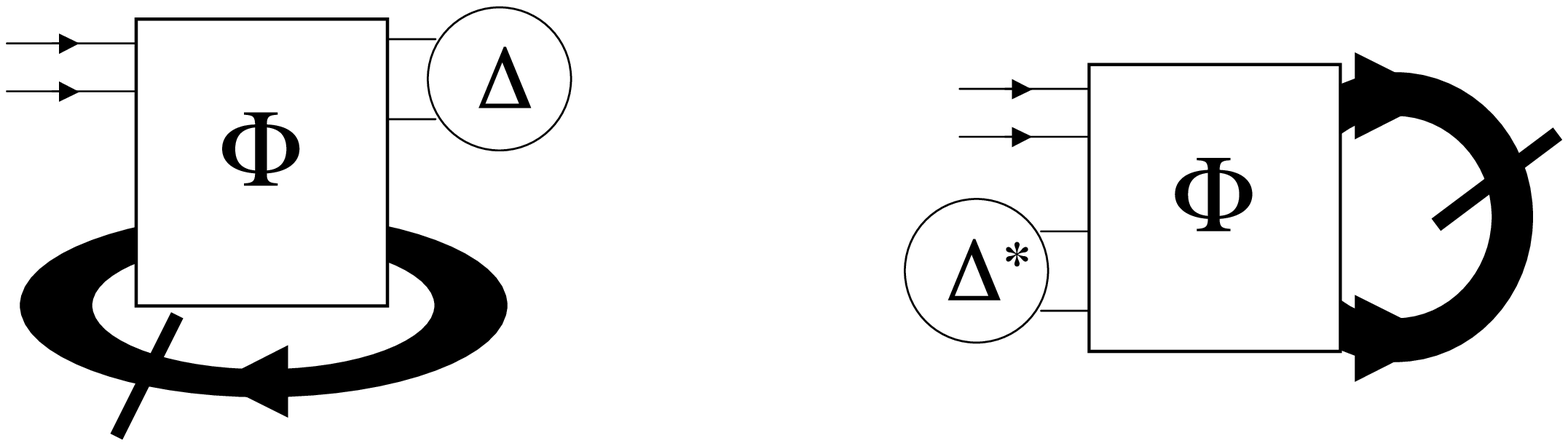}
\caption{Diagrammatic representation of $\Delta_{cm}(p)$. The diagonal component $\Gamma(P)$
of the collective mode propagator is depicted by the heavy line with a single arrow and the off-diagonal one
$\bar{\Gamma}^{a}(P)$ by the heavy line with two arrows. For the two diagrams the slash in the heavy lines
means that the terms which are first order in $|\Delta |^2$ have been removed.}
\label{FigXABCD}
\end{figure}

We will now proceed in a similar way to obtain the mode contribution $\Delta _{cm}(k)$ to the anomalous self-energy $\Delta (k)$. This contribution is obtained by replacing, in the diagrams
which lead to the second term of Eq.(\ref{eqdelta}), any second order product $\Delta \Delta ^{*}$,
$\Delta \Delta$ or $\Delta ^{*} \Delta ^{*}$, by the corresponding collective mode propagator. Accordingly we replace specifically $\Delta \Delta ^{*}$ by the diagonal component $\Gamma(P)$, and $\Delta \Delta$ by the off-diagonal one $\bar{\Gamma}^{a}(P)$. This is depicted diagrammatically in Fig.\ref{FigXABCD}. Just as in Eq.(\ref{eqdelta}) this last term comes with a topological factor $1/2$ to avoid double counting. On the other hand, since the two dimer outgoing lines in the first term play inequivalent roles, such a topological factor does not apply to this first term. It is worth noticing that, since the collective mode propagator depends only on the total momentum-energy four-vector, we do not meet at this stage the problems linked to the $k$-dependence of $\Delta (k)$, discussed in Appendix \ref{appT3}.

Just as in section \ref{anomself}, we should not include the terms of $\Phi(q_1,q_2;p_2, P)$ leading to
reducible contributions to the self-energy. These are easily obtained from the diagrams displayed in
Fig.\ref{FigXXXX} by making the above substitution of $\Delta \Delta ^{*}$ by $\Gamma(P)$. This implies in the
same way the replacement of $\Phi(q_1,q_2;p_2, P)$ by $\Phi'(q_1,q_2;p_2, P)$. However, as in section
\ref{anomself}, it is more convenient to add these reducible terms to both members of the equation for
$\Delta $. In the left-hand side one obtains $[G_0(p)]^{-1}F(p)[G_0(-p)]^{-1}$, while in the right-hand side
one recovers $\Phi$ instead of $\Phi'$. Finally we restrict ourselves to the contribution of the low
frequency collective mode, the higher frequency contributions being expected to be of order $\Delta ^5$
and thus negligible, just as in the preceding section \ref{modnormal}. Accordingly the variable $P$ in
$\Gamma(P)$ and $\bar{\Gamma}^{a}(P)$ is small. This implies that $\Phi(p,-p;P/2,P/2)$, which comes as
a factor of $\Gamma(P)$, can be replaced by $\Phi(p,-p;0,0)$ at the order we are working. Similarly
$\Phi(p,-p;P,0)$, which comes as a factor of $\bar{\Gamma}^{a}(P)$, can also be replaced by $\Phi(p,-p;0,0)$.
After multiplication by $G_0(p)G_0(-p)$ this leads to the equation:
\begin{eqnarray}\label{eqfapp}
F(p)=G_0(p)G_0(-p)\delta_1({\bf p})
+ \,\Delta\, G_0(p)G_0(-p) \Phi (p,-p;0, 0)\left[\frac{1}{2}|\Delta |^2+\sum_{P}{\Gamma}(P)+\frac{1}{2}\sum_{P}{\not \!\bar{\Gamma}}^{a}(P)\right]
\end{eqnarray}
which comes in place of Eq.(\ref{eqf3}). 

In the preceding section \ref{modnormal} we have already calculated $\sum_{P}{\Gamma}(P)$ at the level of Eq.(\ref{eqintx}) and found
$\sum_{P}{\Gamma}(P) =8\pi \sqrt{2 |\mu |}\,m^{-3/2}\,n_{cm}$, where $ n_{cm}$ is defined by 
Eq.(\ref{ncmfin}). In principle the calculation of $\sum_{P}\bar{\Gamma}^{a}(P)$ follows the same line as the
one of $\sum_{P}{\Gamma}(P)$. One performs first the frequency integration by noticing that, for fixed
${\bf P}$, $\bar{\Gamma}^{a}(P)$ has a pole at $\Omega = i \Omega_{cm}({\bf P})$. The corresponding
residue $\bar{Z}^a({\bf P})$ is obtained from the residue $Z({\bf P})$ for $\Gamma(P)$ by comparing
Eq.(\ref{gama}) with Eq.(\ref{gamn}). One has to multiply $Z({\bf P})$ by $\Gamma^a_{\mathrm irr}(P)/\left[T_2^{-1}(-P)-\Gamma_{\mathrm irr}(-P)\right]$, evaluated at the pole, to obtain $\bar{Z}^a({\bf P})$. This gives merely:
\begin{eqnarray}\label{barza}
 \bar{Z}^a({\bf P})=\frac{\Delta}{\Delta^*}\,\frac{4\pi \sqrt{2 |\mu |}}{i\,m^{3/2}}
\frac{\mu _B}{\Omega_{cm}({\bf P})}
\end{eqnarray}
Now if we proceed as in Eq.(\ref{eqintx}) we come to an integral with an ultraviolet divergence, which would invalidate our procedure.

The reason for this problem is that we have not yet consistently removed in our calculation the terms of order zero and one in $|\Delta |^2$, as we have done precedingly. That we should do it was already appearent in Eq.(\ref{eqfapp}) by the notation ${\not \!\bar{\Gamma}}^{a}$ instead of $\bar{\Gamma}^{a}$ (which implies also a multiplication by $\Delta^* /\Delta$). We recall that these terms have to be subtracted out because they have
already been taken into account in our preceding lowest order calculation. However in the above preceding cases, it
turned out that these terms were equal to zero, which made the subtraction transparent. Here this is not the case.
The most efficient way to perform this subtraction is to note that, from Eq.(\ref{gamabar}), $\bar{\Gamma}^{a}(P)$ contains already an explicit factor $\Gamma^a_{\mathrm irr}(P)$ in the numerator, proportional to $|\Delta |^2$, which gives directly the factor $\mu _B$ in Eq.(\ref{barza}). Hence the term of order $|\Delta |^2$ to be subtracted is merely
obtained by taking the other factors to order zero in $|\Delta |^2$, that is by replacing $\Omega_{cm}({\bf P})$ in Eq.(\ref{barza}) by its zeroth order expression ${\bf P}^2/4m$. Actually this argument is a bit too fast. It misses the fact that we may at the same time evaluate $\Phi (p,-p;0, 0)$ in Eq.(\ref{eqfapp}) with the chemical potential taken as $-E_b/2$, even for the $(1/2)|\Delta |^2$ term, as we had done in Eq.(\ref{eqt4}). Hence this whole step is taken up in more details in Appendix \ref{muzero}. We find in this way:
\begin{eqnarray}\label{eqgamaslash}
\sum_{P}{\not \!\bar{\Gamma}}^{a}(P)=-\frac{4\pi \mu _B\sqrt{2 |\mu |}}{m^{3/2}}\int \frac{d{\bf P}}{(2\pi )^3} 
\left[\frac{1}{\Omega_{cm}({\bf P})}-\frac{4m}{{\bf P}^2}\right]
\end{eqnarray}
Making again the change of variables $P=(4m\mu _B)^{1/2}x$, this gives:
\begin{eqnarray}
\sum_{P}{\not \!\bar{\Gamma}}^{a}(P)=-\frac{16\mu _B^{3/2}\sqrt{2 |\mu |}}{\pi } \int_{0}^{\infty}dx\,\left[\frac{x}{\sqrt{x^2+2}}-1\right]=\frac{32}{\pi }\mu _B^{3/2}|\mu |^{1/2} =\frac{24\pi \sqrt{2 |\mu |}}{m^{-3/2}}\,n_{cm}=3\sum_{P}{\Gamma}(P)
\end{eqnarray}
\vspace{10mm}
Hence the last bracket in Eq.(\ref{eqfapp}) can be written $(1/2)|\Delta |^2+(5/2)\sum_{P}{\Gamma}(P)$.

The rest of the argument proceeds just as in section \ref{anomself}, provided $(1/2)|\Delta |^2$ is replaced by the above bracket. This leads to the replacement of Eq.(\ref{mean}) by
\begin{eqnarray}\label{meana}
\frac{1}{a}-\sqrt{2m|\mu|}&=&\frac{m^2 a^2}{4}\, a_M\,\left[\frac{1}{2}|\Delta|^2+\frac{20\pi \sqrt{2 |\mu |}}{m^{3/2}}\,n_{cm}\right]
\end{eqnarray}
We may eliminate $|\Delta |^2$ by making use of the number equation, which reads now from Eq.(\ref{eqdefn2}):
\begin{eqnarray}
n-n_{cm}=\frac{m^2\,|\Delta |^2}{8\pi (2m|\mu |)^{1/2}}
\end{eqnarray}
This gives:
\begin{eqnarray}\label{meanb}
\frac{1}{a}-\sqrt{2m|\mu|}=\pi a^2 a_M\,(2m|\mu |)^{1/2}\left[n+4n_{cm}\right]=\pi a\, a_M\left[n+4n_{cm}\right]
\end{eqnarray}
where in the last equality our replacement of $2m|\mu |$ by $1/a^2$ introduces only errors of order $n^2$ in the
equation of state. To the same order this leads us to:
\begin{eqnarray}
2 \mu = -\frac{1}{ma^2}+\frac{2\pi a_M}{m}\left[n+4n_{cm}\right]
\end{eqnarray}
In the last term we may replace $n_{cm}$ by its expression Eq.(\ref{ncmfin}) $n_{cm}=(4\pi na_M)^{3/2}/3\pi ^2$. This gives the familiar Lee-Huang-Yang result for the chemical potential of a dimer, taking its binding energy as reference:
\begin{eqnarray}
2 \mu +\frac{1}{ma^2} = \frac{4\pi n a_M}{2m}\left[1+\frac{32}{3}\left(\frac{na_M^3}{\pi }\right)^{1/2}\right]
\end{eqnarray}

\section{CONCLUSION}

In this paper we have presented a general theoretical framework to deal with fermionic superfluidity on the BEC
side of the BEC-BCS crossover. Our approach involves no approximation in principle, but in order to obtain explicit
results we have to proceed to an expansion in powers ot the anomalous self-energy, which is basicallly equivalent
to an expansion in powers of the gas parameter $na^3$. We have applied our method to the $T=0$ thermodynamics
and calculated the first terms in the expansion of the chemical potential in powers of the density. 
We have in this way shown explicitely that the mean-field contribution is indeed given 
by the standard expression in terms of the  dimer-dimer scattering length $a_M$. We have then proved that
the Lee-Huang-Yang contribution to the chemical potential retains also its standard expression from elementary
boson superfluid theory provided $a_M$ is used in its expression.  Departures from elementary boson
theory appear at the order $n^2$. Naturally our approach can be used to obtain other physical quantities and
extended for example to finite temperature and dynamical properties. This will be considered elsewhere.

\section{ACKNOWLEDGEMENTS}
"Laboratoire de Physique Statistique de l'Ecole Normale Sup\'erieure" is "associ\'e au Centre National
de la Recherche Scientifique et aux Universit\'es Pierre et Marie Curie-Paris 6 et Paris 7".

\appendix

\section{Expression of $\Sigma^{(2)}$ in terms of $T_3$ in the general case}\label{appT3}

If we note $T_3^{aa}(p,p',p'')$ the sum of all normal diagrams, with one single atom $p$ line going in and out, two entering atom lines $(p',\uparrow$) and $(-p',\downarrow$) corresponding to an ingoing dimer, and two outgoing atom lines $(p'',\uparrow$) and $(-p'',\downarrow$) corresponding to an outgoing dimer, we have:
\begin{eqnarray}\label{eq1appT3}
\Sigma^{(2)}(p)= \sum_{p'\,p"} \Delta ^{*}(p')\Delta (p'') T_3^{aa}(p,p',p'')
\end{eqnarray}
In order to obtain an expression for $T_3^{aa}(p,p',p'')$, one can find in the spirit of Ref.\cite{bkkcl} integral equations for this quantity, and then identify in them a quantity which is precisely $T_3(p_1,p_2;P)$. Another slightly more direct way is to separate, in $T_3^{aa}$, all the diagrams where 0 or 1 dimer lines are present, right at the beginning or right at the end of $T_3^{aa}(p,p',p")$. We do not enter into the details but only give the result, on which the above mentionned separation is clear:
\begin{eqnarray}\label{eq2appT3}
T_3^{aa}(p,p',p'')= -G(-p)\,\delta_{p,p'}\,\delta_{p,p''}+\left[ G(-p') \right]^2\,G(p')\,T_2(p-p')\,\delta_{p',p''}
\nonumber  \\ 
+G(p')G(-p') G(p'')G(-p'')T_2(p-p')T_2(p-p'')T_3(p',p'';p)
\end{eqnarray}
where $\delta_{p,p'}$ is the Kronecker symbol, and $T_2(p)$ is the normal state dimer propagator. One can check, by making use of Eq.~(\ref{eq2appT3}) and of the integral equation \cite{bkkcl} satisfied by $T_3$, that  Eq.~(\ref{eq1appT3}) reduces to Eq.~(\ref{eqsigmn2}) when $\Delta (p)$ does not depend on $p$.

\section{Absence of loops in normal state diagrams}\label{appnoloop}

Since we are at $T=0$ with a negative chemical potential, there is no Fermi sea and no possibility of fermionic holes. When we consider the propagation of a fermionic atom, it has to be created before being destroyed. This corresponds to the fact that the free fermion propagator is retarded. In (imaginary) time and wavevector representation we have for the free atom Green's function:
\begin{eqnarray}
G_0({\bf p},\tau)=-e^{-\xi_{\bf p} \tau} \theta (\tau)
\end{eqnarray}
where $\theta(\tau)$ is the Heaviside function. 

We consider now the contribution coming from any loop in a diagram. For clarity we do not write the wave vectors, which do not play any role in the argument. The loop starts for example at a vertex with time $\tau_0$, with successive interactions at times $\tau_1,\tau_2,...,\tau_n$. This implies in the term corresponding to this diagram the presence of the product $G_0(\tau_0-\tau_1)G_0(\tau_1-\tau_2)....G_0(\tau_n-\tau_0)$. In order to have each term being non zero in this product of retarded Green's function, we need $\tau_0 > \tau_1 > \tau_2 >...>\tau_n > \tau_0$, which is impossible to satisfy.
Hence one of the Green's function in this product will necessarily be zero, and the contribution of any diagram with a loop will also be zero. Note that the argument applies also if there is a single Green's function in the loop \cite{agd}. If we were to carry out calculations in frequency, rather than time, representation the corresponding free atom Green's function is analytical in the upper complex half-plane, and the above result would appear through integration over frequencies with contours closed in the upper complex half-plane. 

We have considered for consistency the imaginary time formalism. Naturally exactly the same arguments apply if we consider the real time representation.

\section{Analytical properties of $T_3(k,k;k+P)$}\label{appT3anal}

We give first an explicit proof of the analytical property of $T_3$ indicated in the text. We start from the integral equation satisfied by $T_3$ with at first essentially the notations of Ref.\cite{bkkcl}:
\begin{eqnarray}\label{3Vertex}
    T_3(p_1,p_2;P)=-G_0(P-p_1-p_2)- \sum_{q}G_0(P-p_1-q)
    G_0(q)\, T_2(P -q)\; T_3(q, p_2; P)
\end{eqnarray}
We follow Ref.\cite{bkkcl} and perform the frequency integration in the second term, making use of the fact that the only singularity in the lower half complex plane comes from $G_0(q)$ and is located at ${\bf q}^2/2m-i0_{+}$. 
We write then explicitely the resulting equation for $p_1=K=\{{\bf K},i\Omega\}$, $p_2=k=\{{\bf k},i\omega \}$,
while $P=\{{\bf P},i\nu \}$ is replaced by $k+P$, keeping in mind that the frequencies $p_1$ and $p_2$ have to be shifted by $\mu $ (and $k+P$ by $3\mu $) and that we work now with imaginary frequencies. This gives:
\begin{eqnarray}\label{}
    T_3(K,k;k+P)\!=\!-\frac{1}{\mu +i(\nu-\Omega)-({\bf P}-{\bf K})^2/2m }-\frac{4\pi }{m^{3/2}}  \!\int \!\!\frac{d{\bf q}}{(2\pi )^3}
   \frac{1}{2\mu +i(\nu-\Omega +\omega)- \!{\bf q}^2/2m\!-\!({\bf k+P-K-q})^2/2m} \nonumber
\end{eqnarray}
\vspace{-3mm}
\begin{eqnarray}\label{3Vertex1}
\hspace{30mm}  \frac{\sqrt{E_b}+\sqrt{({\mathbf P+k-q})^2/4m-(i(\nu+\omega) + 3 \mu -{\bf q}^2/2m)}}
    {i(\nu+\omega) + 3 \mu -{\bf q}^2/2m- ({\mathbf P+k-q})^2/4m+E_b}\;T_3(\{{\bf q},{\bf q}^2/2m-\mu \},k;k+P)
\end{eqnarray}
When we write this equation for $K=k$, we see that the first term, which is the Born contribution, has a pole in the upper complex half-plane, at $\omega =\nu+ i(|\mu |+({\bf P}-{\bf k})^2/2m)$. On the other hand, in the integral, the first factor does not depend on frequency $\omega $. The singularities coming from the dimer term will be in the lower complex half-plane with ${\mathrm Im}\,\omega <-(3|\mu |-E_b)<0$ (since we will have $|\mu | \simeq E_b/2$).
Regarding the $T_3(\{{\bf q},{\bf q}^2/2m - \mu \},k;k+P)$ term, if we write Eq.~(\ref{3Vertex1}) for the variable $K$ evaluated on the shell, i.e. $K=\{{\bf K},{\bf K}^2/2m -\mu \}$, we obtain an integral equation for $T_3(\{{\bf q},{\bf q}^2/2m - \mu \},k;k+P)$. In this equation the Born term does not depend on frequency. The first term in the integral has its poles satisfying ${\mathrm Im}\,\omega <-3|\mu |<0$ and the dimer term has again its singularities in the lower complex half-plane. Hence by a recursive argument (or equivalently by iteration of the equation), we conclude that the singularities of $T_3(\{{\bf q},{\bf q}^2/2m -\mu\},k;k+P)$ are also in the lower complex half-plane. This shows that indeed, except for the Born contribution, $T_3(k,k;k+P)$ is analytical in the upper complex half-plane of the frequency variable $\omega $.

\begin{figure}
\centering
\includegraphics[width=150mm]{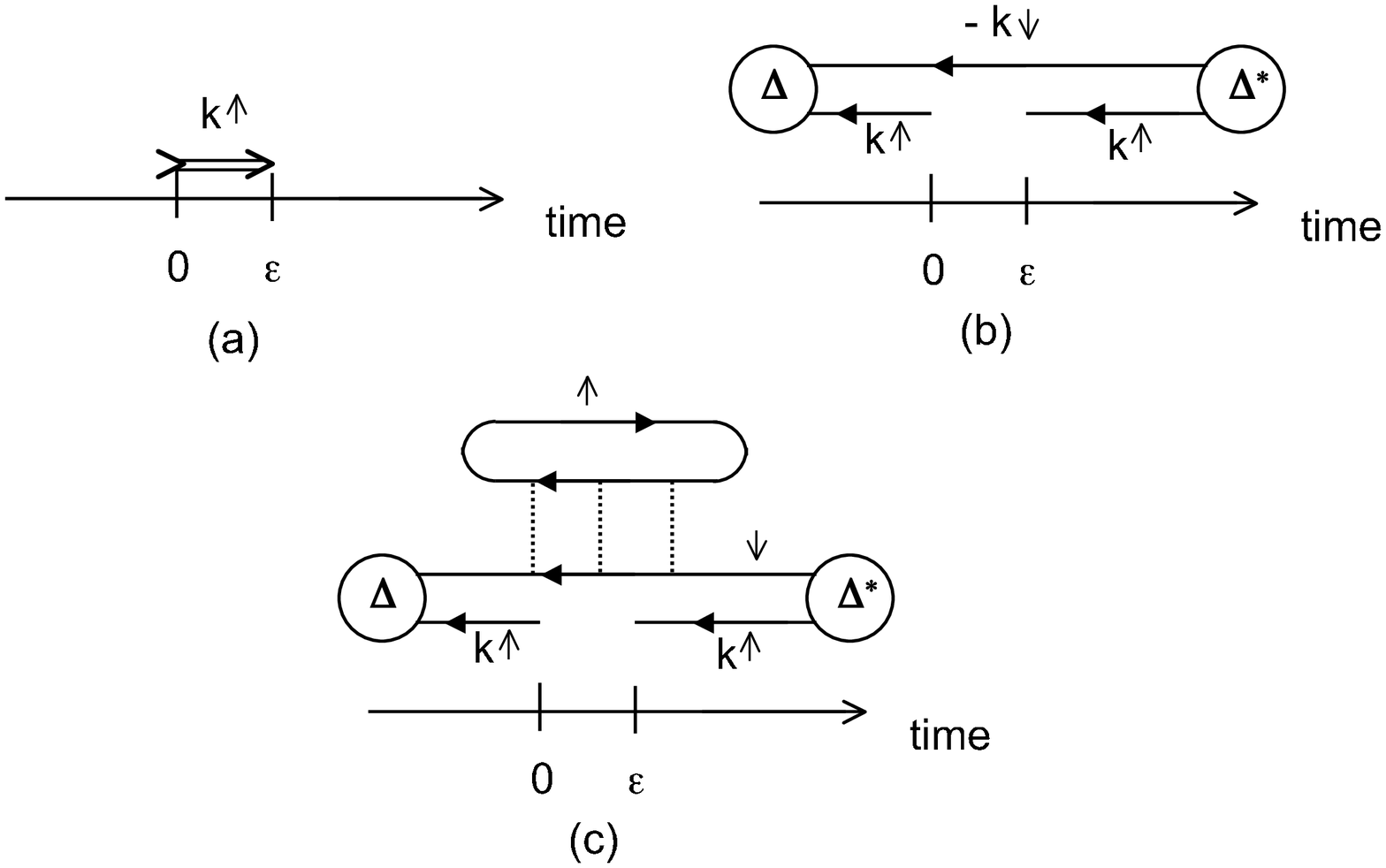}
\vspace{5mm}
\caption{ a) Diagrammatic representation of the propagator $G_{k \uparrow }(-\epsilon)$ 
involved in the calculation of the particle number $n$. We use the standard representation
where the propagator starts at the destruction time and ends at the creation time.
b) Born term in the diagrammatic expansion of the propagator represented in a). c) Example of a diagram
which gives a zero contribution due to the presence of a loop.}
\label{FigAppC}
\end{figure}

Naturally we do not expect such a result to come by chance and we discuss now its physical origin. It is more easily seen in real time, rather than frequency, representation. The particle number Eq.(\ref{eqdefn}) is merely given by:
\begin{eqnarray}
n= \sum_{k} \langle c^{\dagger}_{k \uparrow }(\epsilon ) c_{k \uparrow }(0) \rangle
\end{eqnarray}
with $\epsilon \rightarrow 0_{+}$. The average corresponds to a full propagator $G_{k \uparrow }(-\epsilon)$ in time representation, where the particle is created at time $t=\epsilon $, later than the time $t=0$ at which it is destroyed. In standard \cite{agd} diagrammatic representation (see Fig.\ref{FigAppC} a)), with time going from left to right and the propagator 
starting at the destruction time and ending at the creation time \cite{agd}, 
we have to draw diagrams where the $\uparrow $ particle propagator starts from $t=0 $ and goes globally forward (that is to the right 
in Fig.\ref{FigAppC}) in time, ending up at $t=\epsilon $.
However when we write the perturbation expansion of this propagator, we have to use free particle propagators $G_0$ which are retarded and go only forward in time  (i.e. the arrows go to the left in Fig.\ref{FigAppC} when we use the standard Green's function representation \cite{agd}). Hence the zeroth order in our perturbation expansion is obviously zero. When we go to second order, corresponding to Eq.(\ref{eqdefn1}), we can use $\Delta ^{*}$ which creates, at some given time, a pair of particles $({\bf q}\uparrow ,-{\bf q}\downarrow )$ (this process is instantaneous in time because, at lowest order, $\Delta $ does not depend on frequency). Similarly $\Delta $ annihilates a pair of particles. The simplest combination of these ingredients, giving a non zero result for $n$, is shown in Fig.\ref{FigAppC} b): $\Delta ^{*}$ is located at positive time, and $\Delta $ at negative time, allowing to have $G_0$'s going only to the left. This is just the Born term of $T_3$. However one sees easily that it is not possible to draw any other second order diagrams having the same property shown in Fig.\ref{FigAppC} b). They would necessarily imply $G_0$'s going at some stage backward in time (i.e. with arrows pointing to the right in Fig.\ref{FigAppC}), which is not allowed. Indeed we have to dispose of the $\uparrow $ propagator starting at $t=0 $, and this can only be done by annihilating it with $\Delta $ at some negative time. Similarly the $\uparrow $ propagator ending up at $t=\epsilon $ has to come from $\Delta ^{*}$, acting at some positive time. Then we are left with a $\downarrow $ propagator which can only go from $\Delta ^{*}$ to $\Delta$. Any interaction between the $\uparrow $ and $\downarrow $ propagators is forbidden since this would merely amount to double count processes which are already taken into account in $\Delta ^{*}$ and $\Delta$. In this way we end up with the conclusion that only the Born term leads to a non zero contribution to $n$.

\section{Detailed analysis of the next order contribution to the normal self-energy}\label{normselfa}

In calculating the coefficient of $|\Delta |^4$ in Eq.(\ref{eqcorr1}), it is clear that we can evaluate all the other factors to zeroth order in $\Delta $, since we have already an explicit factor $|\Delta |^4$. This implies in particular to make $2|\mu |=E_b=1/(ma^2)$ in these terms. This leaves us with:
\begin{eqnarray}\label{eqGam}
-\int_{-\infty}^{\infty}\frac{d\Omega }{2\pi }
\,\Gamma(P)=
\int_{-\infty}^{0}\frac{d\omega }{\pi }
\,{\mathrm Im}\left[\Gamma^a_{\mathrm irr}(P){\bar \Gamma}^a_{\mathrm irr}(-P)\,T_2(-P)\right]
|T_2(P)|^2
\end{eqnarray}
with again $P=\{{\bf P},\omega +i\epsilon \}$. Naturally this result could have been obtained more directly,
by a straight expansion of $\Gamma(P)$ in powers of $\Delta $, instead of resumming a whole perturbation series, as we have done in Section \ref{collmod}, and then expanding the result.

We may first apply this formula for the collective mode contribution. In the present case it comes from the factor $T_2(-P)$, which has a pole for $\omega =-{\bf P}^2/4m$. This is indeed the frequency of the Bogoliubov mode, once the frequency is high enough for the zeroth order approximation in $\Delta $ to be valid. Correspondingly we have ${\mathrm Im}\,T_2(-P)=8\pi ^2/(m^2a)\,\delta(\omega +{\bf P}^2/4m)$. On the other hand, we have for this value of the frequency $T_2(P)=(4\pi a/m)[1-\sqrt{1+{\bf P}^2a^2/2}]^{-1}$.
This has to be inserted in Eq.(\ref{eqGam}) and the integration over ${\bf P}$ has still to be performed. However if we concentrate on the low ${\bf P}$ range (where $\Gamma^a_{\mathrm irr}(P){\bar \Gamma}^a_{\mathrm irr}(P) \simeq |\Delta |^4\,T_4^{2}\left(0,0;0\right)/4$), we obtain a divergent integral $\int d{\bf P}/{\bf P}^4 \sim \int d|{\bf P}|/|{\bf P}|^2$. This is  due to the fact that, in this range, our zeroth order handling in $|\Delta |$ is not careful enough. From Eq.(\ref{bog}), we see that the dispersion relation has to be modified for $|{\bf P}|_m \sim (m\mu _B)^{1/2}\sim (na_M)^{1/2} \sim |\Delta |$. If we evaluate the low frequency contribution by putting a lower cut-off $|{\bf P}|_m$ in the integral, we obtain $\int d|{\bf P}|/|{\bf P}|^2 \sim 1/|{\bf P}|_m \sim n^{-1/2} \sim |\Delta |^{-1/2}$ and the overall result is of order $n^{3/2}$. This is just the contribution Eq.(\ref{ncmfin}), where we have obtained the precise coefficient by our careful handling of this low frequency domain. Now, if we define the low momentum region as $|{\bf P}| \le P_0$, where $P_0$ is large compared to
$|{\bf P}|_m$, but small compared to $1/a$, our above treatment of the low frequency region is fully valid in this domain. On the other hand the integral in Eq.(\ref{eqintx}) has already converged at the upper bound $x_0=P_0/(4m\mu _B)^{1/2}$, and we may replace it by $\infty$, just as we have done in Eq.(\ref{eqintx}). Hence the contribution from the domain $|{\bf P}| \le P_0$ is exactly given by Eq.(\ref{ncmfin}). On the other hand, there are no divergences in the domain $|{\bf P}| \ge P_0$ and its contribution is of order
$|\Delta |^4$, and accordingly negligible compared to Eq.(\ref{ncmfin}).

To be complete we have also to verify that there are no other sources of divergence in Eq.(\ref{eqGam}), analogous to what we have found at low momentum. Since there is no singularity arising at finite frequency or momentum, the only possibility comes from the large frequency $|\omega |>\Omega_c$, large momentum $|{\bf P}|>P_c$ region. In order to evaluate
the contribution in this domain, it is more convenient to go back to the imaginary axis contour by performing
backward the first step of Eq.(\ref{eqcorr1}) and write this overall
contribution as:
\begin{eqnarray}
- \int_{|{\bf P}|>P_c} \frac{d{\bf P}}{(2\pi )^3} \int_{|\Omega |>\Omega_c}\frac{d\Omega }{2\pi }
\;\Gamma^a_{\mathrm irr}(P){\bar \Gamma}^a_{\mathrm irr}(-P)\,T^2_2(-P)
T_2(P)
\end{eqnarray}
The scattering amplitudes $T_4\left(P,0;0\right)$ and $T_4\left(0,-P;0\right)$, introduced by $\Gamma^a_{\mathrm irr}(P)$ and ${\bar \Gamma}^a_{\mathrm irr}(-P)$, may be
shown to be equal and are dominated as expected in this range by the Born contribution. They are real on this imaginary frequency axis. The Born term has been obtained explicitely in \cite{bkkcl}, and $T_4\left(P,0;0\right)$ is proportional to $1/(|{\bf P}||\Omega |)$ for large $|{\bf P}|$ and $|\Omega |$. From Eq.(\ref{2Vertex}) there is in the modulus of the integrand a factor $1/\left[\Omega^2+({\bf P}^2/4m)^2 \right]^{3/4}$ coming from $T^2_2(-P)\,T_2(P)$. Then the overall integral is easily seen to be convergent for large $|{\bf P}|$ and $|\Omega |$, which completes our proof.

\section{Details on the contribution of the collective mode to the anomalous self-energy}\label{muzero}

\begin{figure}
\centering
\includegraphics[width=80mm]{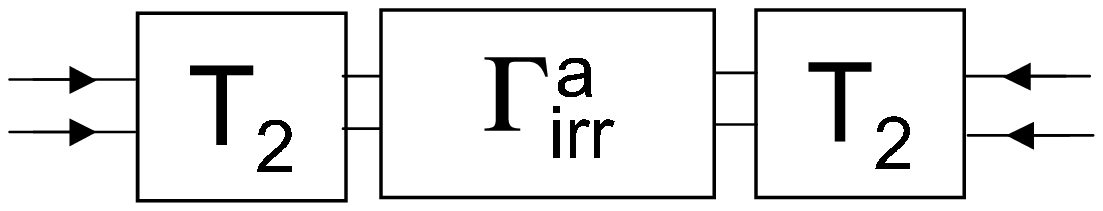}
\caption{Diagrammatic representation of the difference between ${\not \!\bar{\Gamma}}^{a}(P)$ and $\bar{\Gamma}^{a}(P)$}
\label{FigDiagsub}
\end{figure}

Since there is no zeroth order term in powers of $|\Delta |$ in $\bar{\Gamma}^{a}(P)$, we have only to
remove the term proportional to $|\Delta |^2$ to avoid double counting, and go in this way from $\bar{\Gamma}^{a}(P)$ to ${\not \!\bar{\Gamma}}^{a}(P)$. The corresponding diagram in $F(p)$ is shown in Fig.\ref{FigDiagsub}. This leads to:
\begin{eqnarray}
{\not \!\bar{\Gamma}}^{a}(P)-\bar{\Gamma}^{a}(P)=-\Gamma^a_{\mathrm irr}(P)T_2(P)T_2(-P)
\end{eqnarray}
where $\Gamma^a_{\mathrm irr}(P)$ is given by Eq.(\ref{gammirra}). Let us first replace the dimer propagator $T_2(P)$, given by Eq.(\ref{2Vertex}), by its value $T^0_2(P)$ where $\mu$ is replaced by its zeroth order value $-E_b/2$. When we consider the contribution from the low frequency and momentum contribution, we may replace $\Gamma^a_{\mathrm irr}(P)$ by $\Gamma^a_{\mathrm irr}(0)$, and the pole of $T^0_2(-P)$ at $\Omega = i {\bf P}^2/(4m)$ gives:
\begin{eqnarray}
-\int_{-\infty}^{\infty}\frac{d\Omega }{2\pi } \left[-\Gamma^a_{\mathrm irr}(P)T^0_2(P)T^0_2(-P)\right]=
\frac{4\pi \mu _B\sqrt{E_b}}{m^{3/2}} 
\frac{4m}{{\bf P}^2}
\end{eqnarray}
This is indeed the second term in the right-hand side of Eq.(\ref{eqgamaslash}) since in this last equation,
which turns out to be of order $|\Delta |^3$, we have to make consistently $2|\mu |=E_b$ (and similarly we have to take $\mu _B=4\pi \sqrt{E_b}\Gamma_{\mathrm irr}(0)/m^{3/2}$, with $\Gamma_{\mathrm irr}(0)=2(\Delta ^*/\Delta )\Gamma^a_{\mathrm irr}(0)$).

We are now left with the evaluation of the difference between our original expression and the one we have
just considered, namely:
\begin{eqnarray}
-\Gamma^a_{\mathrm irr}(P)\left[T_2(P)T_2(-P)-T^0_2(P)T^0_2(-P)\right]
\end{eqnarray}
which gives at the level of Eq.(\ref{eqgap3}), after insertion at the level of Eq.(\ref{eqfapp}) and summation over $p$ and $P$:
\begin{eqnarray}\label{eqabc}
-\frac{1}{4}\Delta^*\,\Delta^{2} 
\sum_{P}\;T_4\left(0,P;0\right)
\left[T_2(P)T_2(-P)-T^0_2(P)T^0_2(-P)\right] T_4(P,0;0) 
\end{eqnarray}
where we have used Eq.(\ref{gammirra}) for $\Gamma^a_{\mathrm irr}(P)$ and the relation (see Ref. \cite
{bkkcl}) between $T_4$ and $\Phi$, namely $T_4(0,P;0)=\sum_{p} G_0(-p)G_0(p)\Phi(-p,p;P,0)$.

We show now that this difference is just what is required to be allowed to replace $\frac{1}{2}  \,\Delta\, |\Delta |^2T_4(0,0;\{{\bf 0},2\mu \})=\frac{1}{2}  \,\Delta\, |\Delta |^2\sum_{p} G_0(-p)G_0(p)\Phi(-p,p;0,0)$ in Eq.(\ref{eqt4}) by $\frac{1}{2}  \,\Delta\, |\Delta |^2T_4(0,0;\{{\bf 0},-E_b\})$, which is directly
related to the dimer-dimer scattering scattering length $a_M$ by Eq.(\ref{eqaM}). Indeed it has been shown
in \cite{bkkcl} that scattering amplitude $T_4(p_1,p_2;P)$ satisfies quite generally an integral equation 
analogous to the standard integral equation for elementary bosons, namely:
\begin{eqnarray}
 T_4(p_1,p_2;P)=\Delta_4(p_1,p_2;P)+\frac{1}{2}\sum\limits_q
    \Delta_4(p_1,q;P)T_2(P+q)T_2(P-q)T_4(q,p_2;P)
\end{eqnarray}
where $\Delta_4(p_1,p_2;P)$ is the sum of all irreducible diagrams with respect to cutting two dimer propagators, which is itself obtained from an integral equation with explicit kernel \cite{bkkcl}. Formally
this equation can be rewritten in terms of operators as:
\begin{eqnarray}\label{eqform}
\Delta_4 ^{-1}=T_4^{-1}+D
\end{eqnarray}
where we have introduced the diagonal operator $D(q;P) \equiv (1/2)T_2(P+q)T_2(P-q)$ (the factor $1/2$ is again topological).
We consider now this equation for $p_1=p_2=0$ and ${\bf P}={\bf 0}$. With respect to the energy variable
corresponding to $P$, we take the two values $2\mu $ and $-E_b$ (we note the operators with this last value
with a superscript $0$). The crucial point is that the difference between $\Delta _4$ and $\Delta _4^0$ is
of order $2\mu+E_b$, that is the expansion of $\Delta _4$ in powers of $2\mu+E_b$ is regular. Such a
difference is negligible at the order we are working at. The regularity of the expansion can be
seen from the integral equation (Eq.(B2) of Ref.\cite{bkkcl}) satisfied by $\Delta _4$, which is completely
analogous to the one satisfied by $\Phi$ (Eq.(9) of Ref.\cite{bkkcl}), except that the last term (with two
dimer propagators) in the equation for $\Phi$ is not present for $\Delta _4$. This integral equation can be transformed in the
same way, leading to more explicit expressions (see Eq.(11) and Eq.(13) of Ref.\cite{bkkcl}). On these
expressions one checks easily that, for $\mu $ in the vicinity of $-E_b/2$, the propagator $T_2$ is
evaluated for values of its argument which are never in the vicinity of its pole. Hence one can proceed
to a regular expansion. By contrast one sees also on the equations for $\Phi$ that, for the term containing two
dimer propagators the argument does reach the pole value, which makes the expansion singular. This
explains why we can not argue that the difference between $T _4(0,0;\{{\bf 0},2\mu \})$ and
$T_4(0,0;\{{\bf 0},-E_b\})$ is of order $2\mu+E_b$ and negligible. The equation for $P=\{{\bf P},-E_b\}$
reads:
\begin{eqnarray}\label{eqform0}
\left[\Delta_4^0\right]^{-1}=\left[T_4^0\right]^{-1}+D^0
\end{eqnarray}
Writing the difference with Eq.(\ref{eqform}) and taking into account $\Delta _4 \simeq \Delta _4^0$, we have:
\begin{eqnarray}\label{eqform00}
\left[T_4^0\right]^{-1}-T_4^{-1}=D-D_0
\end{eqnarray}
which can be rewritten explicitely, after premultiplication by $T_{4,0}$ and postmultiplication by $T_{4}$, as:
\begin{eqnarray}
T_4(0,0;\{{\bf 0},2\mu \})-T_4(0,0;\{{\bf 0},-E_b\})=\frac{1}{2}\sum\limits_q
T_4(0,q;\{{\bf 0},-E_b\})\left[T_2(q)T_2(-q)-T^{0}_2(q)T^{0}_2(-q)\right]T_4(q,0;\{{\bf 0},-E_b\})
\end{eqnarray}
where in the last term we have been allowed to replace $2\mu $ by $E_b$ at the order we are
working. The same substitution is allowed in Eq.(\ref{eqabc}), which completes our proof.

Let us finally note that, with respect to this part, our procedure has obvious strong analogies with the work
of Beliaev \cite{bel}. Indeed, as soon as we have established the non trivial result that, at our level
of approximation, the structure of $\Delta _4$ can be omitted which makes it analogous to
an effective interaction, our dimers behave as elementary bosons and we are basically back to the situation
investigated by Beliaev.


\vspace{10mm}

\begin{references}
\bibitem{gps}For a very recent review, see S. Giorgini, L. P. Pitaevskii and S. Stringari, arXiv:0706.3360
and to be published in Rev. Mod. Phys.
\bibitem{bcs}J. Bardeen, L. N. Cooper and J. R. Schrieffer, Phys. Rev. {\bf 108}, 1175 (1957).
\bibitem{pitcoop}L. P. Pitaevskii, Zh. Eksp. Teor. Phys. {\bf 37}, 1794 (1959) [Sov. Phys. JETP {\bf 10}, 1267 (1960)];
L. N. Cooper, R. L. Mills and A. M. Sessler, Phys. Rev. {\bf 114}, 1377 (1959).
\bibitem{mosblat}S. A. Moskalenko, Fiz. Tverd. Tela {\bf 4}, 276 (1962) [Sov. Phys. Solid State {\bf 4}, 199 (1962)];
J. M. Blatt, K. W. B\"{o}er and W. Brandt, Phys. Rev. {\bf 126}, 1692 (1962).
\bibitem{kk}L. V. Keldysh and A. N. Kozlov, Zh. Eksp. Teor. Phys. {\bf 54}, 978 (1968)
[Sov. Phys. JETP {\bf 27}, 521 (1968)]
\bibitem{popov} V. N. Popov, Zh. Eksp. Teor. Phys. {\bf 50}, 1550 (1966), [Sov. Phys. JETP {\bf 23}, 1034 (1966)].
\bibitem{eagles}D. M. Eagles, Phys. Rev. {\bf 186}, 456 (1969); D.M. Eagles, R.J. Tainsh, C. Andrikidis, Physica C {\bf 157}, 48 (1989).
\bibitem{leg}A. J. Leggett, J. Phys. (Paris), Colloq. {\bf 41}, C7-19 (1980); in \emph{Modern Trends in the Theory of
Condensed Matter}, edited by A. Pekalski and J. Przystawa (Springer, Berlin)
\bibitem{nsr}P. Nozi\`eres and S. Schmitt-Rink, J. Low Temp. Phys. {\bf 59}, 195 (1985). 
\bibitem{sdm}C.A.R. S\' a de Melo, M. Randeria and J.R. Engelbrecht, Phys. Rev. Lett. {\bf
71}, 3202 (1993).
\bibitem{rcfesh}R. Combescot, Phys. Rev. Lett. {\bf 91}, 120401 (2003).
\bibitem{haus}R. Haussmann, Z. Phys. B {\bf 91}, 291(1993).
\bibitem{pieristrinati}P. Pieri and G. C. Strinati, Phys. Rev. Lett. {\bf 91}, 030401 (2003).
\bibitem{pierist}P. Pieri and G. C. Strinati, Phys. Rev B, {\bf 61}, 15370 (2000).
\bibitem{petrov}D. S. Petrov, C. Salomon, and G.V. Shlyapnikov, Phys. Rev. Lett. {\bf 93}, 090404 (2004).
\bibitem{bkkcl}I.V. Brodsky, A.V. Klaptsov, M.Yu. Kagan, R. Combescot and X. Leyronas, J.E.T.P. Letters {\bf 82}, 273 (2005) and Phys. Rev. A {\bf 73}, 032724 (2006).
\bibitem{xlrc}X. Leyronas and R. Combescot, Phys. Rev. Lett. {\bf 99}, 170402 (2007).
\bibitem{hbar}We set $\hbar=1$ throughout the paper.
\bibitem{lhy} T. D. Lee and C. N. Yang, Phys. Rev. {\bf 105}, 1119 (1957);
T. D. Lee, K. Huang and C. N. Yang, Phys. Rev. {\bf 106}, 1135 (1957).
\bibitem{bel}S. T. Beliaev, Sov. Phys. JETP {\bf 7}, 289 (1958).
\bibitem{sandro}S. Stringari, Europhys. Lett. {\bf 65}, 749 (2004).
\bibitem{acls}G. E. Astrakharchik, R. Combescot, X. Leyronas and S. Stringari, Phys. Rev. Lett. {\bf 95}, 030404 (2005).
\bibitem{abcg}G. E. Astrakharchik, J. Boronat, J. Casulleras, and S. Giorgini, Phys. Rev. Lett. {\bf 93}, 200404 (2004).
\bibitem{altm}A. Altmeyer, S. Riedl, C. Kohstall, M. J. Wright, R. Geursen, M. Bartenstein, C. Chin, J. Hecker Denschlag, and R. Grimm, Phys. Rev. Lett. {\bf 98}, 040401 (2007).
\bibitem{agd}For an introduction to many-body techniques and superconductivity, see for example 
A. A. Abrikosov, L. P. Gorkov and I. E. Dzyaloshinski, {\it Methods of Quantum Field Theory in Statistical Physics} (Dover, New York, 1975) and A. L. Fetter and J. D. Walecka, \emph{Quantum Theory of Many-Particle Systems} (McGraw-Hill, New York, 1971).
\bibitem{vw}D. Vollhardt and P. W\"olfle, {\it The Superfluid phases of Helium 3} (Taylor and Francis, 1990) and references therein.
\bibitem{galit}V. M. Galitskii, Sov. Phys. JETP {\bf 7}, 104(1958).
\bibitem{ps}For an introduction and references, see L. Pitaevskii and S. Stringari, \emph{Bose-Einstein Condensation} (Oxford, 2003).
\end{references}
\end{document}